\documentclass{article}

\usepackage{arxiv}

\usepackage[utf8]{inputenc}
\usepackage[T1]{fontenc}
\usepackage{hyperref}
\usepackage{url}
\usepackage{booktabs}
\usepackage{amsfonts}
\usepackage{amsmath,amssymb}
\usepackage{nicefrac}
\usepackage{microtype}
\usepackage{graphicx}
\usepackage[numbers,sort&compress]{natbib}
\usepackage{algorithm}
\usepackage{multirow}
\usepackage{array}
\usepackage{lscape}
\usepackage{xcolor}
\definecolor{barcol}{HTML}{4C78A8}       
\usepackage{colortbl}                    
\usepackage{eso-pic}     
\usepackage{calc}        
\usepackage[framemethod=tikz]{mdframed}  
\usepackage{float}                       
\usepackage{subcaption}                  
\usepackage{listings}                    
\usepackage{adjustbox}
\usepackage{enumitem}
\usepackage{xspace}
\usepackage{tcolorbox}
\tcbuselibrary{listings,breakable}
\usepackage{algpseudocode}
\usepackage{wrapfig}
\usepackage[normalem]{ulem}

\usepackage{fontawesome5}
\usepackage{bxcoloremoji}

\definecolor{claude46}{HTML}{EEEDFE}
\definecolor{claude46text}{HTML}{534AB7}
\definecolor{gpt5m}{HTML}{E1F5EE}
\definecolor{gpt5mtext}{HTML}{0F6E56}
\definecolor{gpt5}{HTML}{E6F1FB}
\definecolor{gpt5text}{HTML}{185FA5}
\definecolor{claude45}{HTML}{FAEEDA}
\definecolor{claude45text}{HTML}{854F0B}
\definecolor{bestrow}{HTML}{F6F5F3}
\definecolor{bestval}{HTML}{1D9E75}
\definecolor{dimtext}{HTML}{999999}
\definecolor{gpt4o}{HTML}{FDE8E8}
\definecolor{gpt4otext}{HTML}{B03A2E}
\definecolor{claude37}{HTML}{E0F4F4}
\definecolor{claude37text}{HTML}{0E7490}
\definecolor{gemini25p}{HTML}{EEF2FF}
\definecolor{gemini25ptext}{HTML}{3730A3}
\definecolor{amazonq}{HTML}{FFF3E0}
\definecolor{amazonqtext}{HTML}{E65100}
\definecolor{lightgray}{gray}{0.9}

\newcolumntype{R}{>{\raggedleft\arraybackslash}p{1.8cm}}
\newcommand{\cellbest}[1]{\cellcolor{bestrow}#1}
\newcommand{\cellbestval}[1]{\cellcolor{bestrow}\textcolor{bestval}{\textbf{#1}}}

\newcommand{\mypipeline}{DPIAgent\xspace}

\definecolor{abstractbg}{RGB}{232, 240, 253}  
\definecolor{abstractcode}{RGB}{180, 50, 70}  
\definecolor{abstractlink}{RGB}{20, 90, 170}  
\definecolor{msblue}{HTML}{0078D4}
\definecolor{msred}{HTML}{F25022}
\definecolor{msgreen}{HTML}{7FBA00}
\definecolor{msyellow}{HTML}{FFB900}

\definecolor{revision}{RGB}{200,0,0}
\definecolor{revisionb}{RGB}{0,0,200}
\newif\ifshowrevisions
\showrevisionsfalse
\newcommand{\revision}[1]{\ifshowrevisions\textcolor{revision}{#1}\else#1\fi}

\renewenvironment{abstract}{}{}

\providecommand{\keywords}[1]{}
\renewcommand{\keywords}[1]{}

\InputIfFileExists{gallery_runs/table1_cells.real.tex}{}{}
\InputIfFileExists{ablation_runs/abl_cells.tex}{}{}

\title{DPIAgent: Divide, Protocol, Isolate for Agentic Reproduction Test Generation}

\makeatletter
\renewcommand{\@toptitlebar}{%
  \noindent\hspace*{-0pt}\rule{\dimexpr\textwidth+0pt\relax}{2pt}%
  \vskip 0.3in
  \vskip -\parskip
}
\renewcommand{\@maketitle}{%
  \vbox{%
    \hsize\textwidth
    \linewidth\hsize
    \vskip 0.1in
    \@toptitlebar
    \centering
    {\LARGE\sffamily\bfseries \@title\par}%
    \vskip 0.15in
    \textsc{\undertitle}%
    \def\And{%
      \end{tabular}\hfil\linebreak[0]\hfil%
      \begin{tabular}[t]{c}\ignorespaces%
    }
    \def\AND{%
      \end{tabular}\hfil\linebreak[4]\hfil%
      \begin{tabular}[t]{c}\ignorespaces%
    }
    \begin{tabular}[t]{c}\ignorespaces\@author\end{tabular}%
    \vskip 0.25in \@minus 0.05in \center{\@date}\vskip 0.15in
  }
}
\makeatother

\author{%
Hao Liu\textsuperscript{1*\dag} \quad Steven Liu\textsuperscript{1*\dag} \quad  Xin Zhang\textsuperscript{1*\ddag} \quad Jane Luo\textsuperscript{1\dag} \quad Yu Kang\textsuperscript{1} \quad Jie Wu\textsuperscript{1,2\dag} \\
Fangkai Yang\textsuperscript{1} \quad Yangyu Huang\textsuperscript{1} \quad Pengfei Gao\textsuperscript{1} \quad Scarlett Li\textsuperscript{1} \quad Yan Lu\textsuperscript{1}\\[2pt]
{\small\color{black!55}\textsuperscript{1}Microsoft \quad \textsuperscript{2}Tsinghua University}%
}

\renewcommand{\headeright}{}
\providecommand{\undertitle}{}
\renewcommand{\undertitle}{}

\begin{document}
\newcommand{\monthyear}{\ifcase\month\or
  January\or February\or March\or April\or May\or June\or
  July\or August\or September\or October\or November\or December\fi
  \space\number\year}
\newlength{\dateboxwidth}
\settowidth{\dateboxwidth}{\normalsize\monthyear}
\AddToShipoutPictureBG*{%
  \AtPageUpperLeft{%
    \put(64pt,-108pt){\includegraphics[height=96pt]{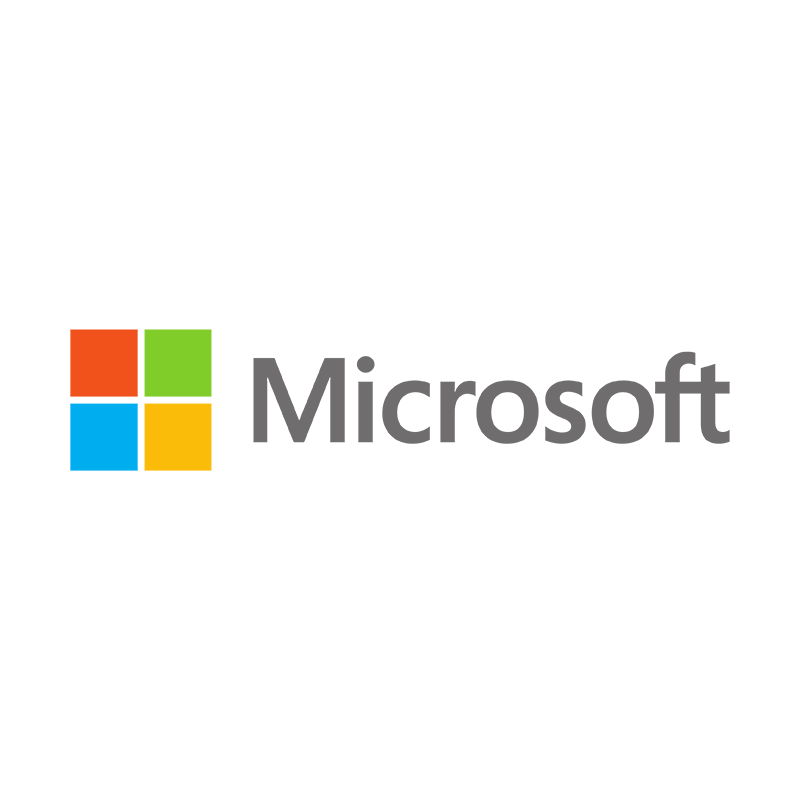}}%
    \put(\dimexpr\paperwidth-72pt-\dateboxwidth\relax,-72pt){\normalsize\monthyear}%
  }%
}
\date{}
\maketitle

\renewcommand{\thefootnote}{}
\makeatletter
\long\def\@makefntext#1{\noindent #1}
\makeatother
\hypersetup{hidelinks}
\footnotetext{\textsuperscript{*}Equal contribution. \quad \textsuperscript{\ddag}Corresponding author. \\ \textsuperscript{\dag}Work done during internships at Microsoft Research Asia.}
\addtocounter{footnote}{-1}

\begin{abstract}
\begin{mdframed}[backgroundcolor=abstractbg,
                 linewidth=0pt,
                 roundcorner=12pt,
                 leftmargin=12pt,
                 rightmargin=12pt,
                 innerleftmargin=16pt,
                 innerrightmargin=16pt,
                 innertopmargin=16pt,
                 innerbottommargin=16pt,
                 skipabove=8pt,
                 skipbelow=8pt]
\begingroup
\let\origtexttt\texttt
\renewcommand{\texttt}[1]{{\color{abstractcode}\origtexttt{#1}}}
\hypersetup{urlcolor=abstractlink}
\hypersetup{hidelinks}

Reproduction test generation, producing a failing-then-passing test that captures a reported bug, is a critical step in automated software engineering. Existing agentic methods treat this as a monolithic loop, despite the task inherently comprising two subtasks of distinct nature: diagnosing the root cause and writing a fail-to-pass test. Without explicit separation, the agent faces a compound objective with underspecified intermediate goals, leading to goal drift. We propose \mypipeline{}, a structured agentic framework built on three principles, Divide, Protocol, Isolate (DPI), that mitigates compound-objective ambiguity and goal drift: it Divides the task into single-objective phases of defect exploration and test generation; enforces a handoff Protocol that records the diagnosis and test plan, preventing context loss; and Isolates each phase's action space by tailoring the toolset to its task, preventing irrelevant tools from misleading execution. On SWT-Bench Verified, \mypipeline{} outperforms seven baselines across three backbone LLMs. With DPI alone it reaches 81.76\% success rate on GPT-5, the highest reported among open-source methods, gaining up to 11.88 points over the strongest baseline on GPT-5-Mini; adding test selection further raises it to 86.17\%. Our analysis shows that architectural structure and backbone capability are complementary axes rather than substitutes, demonstrating DPI's generalizability across model classes.

\coloremojicode{1F517}~ \textbf{Project}: ~\href{https://neverwinhao.github.io/DPIAgent/}{\color{abstractlink}https://neverwinhao.github.io/DPIAgent/}

\coloremojicode{1F4E7}~ \textbf{Correspondence}: \href{haoliu.ai@outlook.com}{haoliu.ai@outlook.com};  \href{xinzhang3@microsoft.com}{xinzhang3@microsoft.com}

\endgroup
\end{mdframed}
\end{abstract}


\section{Introduction}
\label{sec:intro}

Large Language Models (LLMs)~\citep{hurst2024gpt,yang2025qwen3,comanici2025gemini} have demonstrated strong capabilities across various domains and are increasingly applied to code generation~\citep{codex,qwen2.5-coder,rpg} and software engineering automation. A particularly important yet underexplored task is reproduction test generation: given a bug report, the goal is to produce a test that fails on the buggy code and passes after the fix. Despite its central role in the software development lifecycle, reproduction tests are rarely available in the test suite when an issue is first reported~\citep{Libro,swtbench,testexplora}. And developers widely consider writing tests a tedious and time-consuming task~\citep{straubinger2023survey}.

Existing approaches to reproduction test generation fall into two paradigms.
The first follows a \textbf{workflow design}~\citep{otter,e-otter,issue2test}, where the task is decomposed into fixed pipeline stages such as localization, planning, and generation, with predefined control flow orchestrating LLM calls at each stage. The second adopts an \textbf{agent paradigm}~\citep{sweagent,openhands,traeagent}, where the LLM is equipped with tools and dynamically decides which actions to take and when to terminate, without hardcoded stage transitions. The agent paradigm offers key advantages: it allows the model to pivot based on runtime observations and maintains reasoning continuity throughout the task, achieving strong performance on repository-level benchmarks~\citep{swebench,swtbench,swebenchlive}.
Recent leading systems for repository-scale code tasks are predominantly agent-based; we therefore build our method in the agent paradigm.

\begin{figure}[!ht]
\centering
\includegraphics[width=\textwidth]{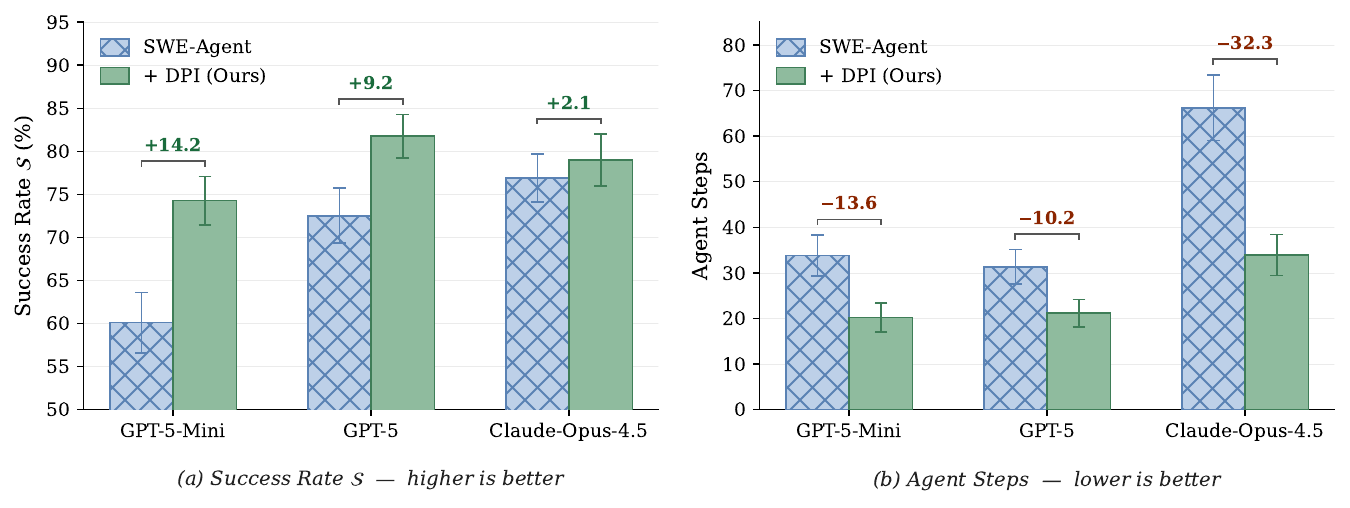}
\caption{
  \textbf{Isolating DPI's contribution.}
 Adding DPI to SWE-Agent yields consistent gains in success rate and reduces agent steps across three backbones. GPT-5-Mini+DPI surpasses GPT-5 SWE-Agent, indicating structural intervention can compensate for model scale.
}
\label{fig:teaser}
\vspace{-10pt}
\end{figure}

However, while the agent paradigm offers flexibility, existing approaches treat reproduction test generation as a monolithic task, mixing exploration and test writing in an undifferentiated loop.
In reality, this task inherently comprises two subtasks of distinct nature: understanding the root cause through code exploration, and producing a test satisfying the fail-to-pass criterion.
Without explicit separation, the agent faces a compound objective with underspecified intermediate goals, which has been shown to systematically induce goal drift~\citep{goaldrift} and is identified as a primary failure mode for software engineering agents~\citep{ambig-swe}.
In practice, developers naturally follow a two-stage process of first localizing the root cause and then writing the test~\citep{BRTAgent}, suggesting that explicit decomposition aligns with the inherent structure of the task.
Moreover, developers locate relevant code through their understanding of code structure and dependencies, whereas existing agents lack such awareness and rely on basic text-matching tools, demanding extensive back-and-forth in large repositories.

Motivated by these observations, we propose \mypipeline{}, an agent design that imposes structure \emph{only at phase boundaries} while preserving a full agent loop within each phase.
\mypipeline{} follows three principles---\textbf{D}ivide, \textbf{P}rotocol, \textbf{I}solate (\textbf{DPI}).
\textbf{(D)ivide:} \mypipeline{} replaces the monolithic loop with two sequential phases, defect exploration and test generation. Each phase has a single, well-defined objective, eliminating the compound-objective ambiguity of monolithic agents.
\textbf{(P)rotocol:} Agents cannot transition silently: it must explicitly write down its diagnosis and test plan before entering the next phase. This forces an explicit handoff so no context is lost between exploration and test writing.
\textbf{(I)solate:} Agents often struggle with tool distraction when exposed to an overly broad action space. To maintain focus, we tailor each phase's toolset strictly to its task, preventing irrelevant tools from misleading the agent's execution.
In addition, we adopt a test selection step from e-Otter++~\citep{e-otter} that turns variance across diversified candidate runs into improved final output quality.

We evaluate \mypipeline{} on SWT-Bench Verified across three backbone LLMs (GPT-5-Mini~\citep{gpt5-system-card}, GPT-5~\citep{gpt5-system-card}, and Claude-Opus-4.5\footnote{\url{https://www.anthropic.com/news/claude-opus-4-5}}), comparing against seven baselines spanning diverse agent architectures. 
Results show that \mypipeline{} consistently outperforms all baselines across all backbones, reaching 81.76\% with DPI alone on GPT-5 (the highest among open-source methods on this benchmark) and 86.17\% with test selection.
Isolating DPI's effect from confounds in baseline architectures, we observe consistent gains across all backbones in both pass rate and step efficiency (Fig.~\ref{fig:teaser}). These gains diminish but persist as model capability grows, indicating that structural design and model scale are complementary axes rather than substitutes.
Ablation studies verify the effectiveness of DPI's design components.
\revision{First-edit quality analysis shows that DPI consistently improves pre-feedback test readiness across all backbones, especially benefiting weaker models that rarely recover from incomplete initial test designs.}

Our contributions can be summarized as follows:
\begin{itemize}[leftmargin=*, nolistsep]
\setlength{\itemsep}{1mm}
\item We propose \mypipeline{}, a structured agentic framework that mitigates compound-objective ambiguity and goal drift in composite tasks via the \textbf{DPI} principles. It divides monolithic tasks into single-objective phases, enforces a structured handoff protocol to preserve context, and isolates phase-specific toolsets to prevent action-space distraction.

\item On SWT-Bench Verified, \mypipeline{} outperforms seven baselines across three backbone LLMs. With DPI alone, \mypipeline{} reaches \textbf{81.76\%} success rate on GPT-5, the highest reported among open-source methods, and gains up to 11.88 points over the strongest baseline on GPT-5-Mini. Adding test selection further raises the result to 86.17\% on GPT-5.

\item Our analysis reveals that \textbf{architectural structure and backbone capability are complementary axes rather than substitutes}. \mypipeline{} yields the largest absolute gain on the weakest model, 11.88 points on GPT-5-Mini, demonstrating DPI's generalizability across model classes. It further reaches peak pass rates in 3$\times$ fewer steps, uniquely resolves 18+ issues, and improves function-level localization by 25+ Acc@5 points.
\end{itemize}

\section{Related Work}
\paragraph{Workflow-Based Test Generation} One line of work decomposes test generation into fixed pipeline stages with predefined control flow. 
Early methods focus on post-hoc selection or assertion manipulation without explicit localization~\citep{Libro,assertflip}. Subsequent work introduces localization as a prerequisite: Issue2Test~\citep{issue2test} and Auto-TDD~\citep{ahmed2024tdd} perform root-cause analysis before generation, SWE-Tester~\citep{swetester} trains open-source LLMs with BM25-based retrieval, and the Otter series~\citep{otter,e-otter} progressively enriches the pipeline with self-reflective planning, execution-augmented repair, and surrogate patch-based selection.
While these methods benefit from explicit stage structure, they rely on rigid control flow that cannot adapt to runtime observations.

\paragraph{Agent-Based Test Generation}
A second line applies general-purpose LLM agents that dynamically decide actions based on observations.
Representative systems include SWE-Agent+~\citep{swtbench}, OpenHands~\citep{openhands}, and TraeAgent~\citep{traeagent}, which differ in tool design and candidate selection but all treat test generation as a monolithic task, mixing exploration and writing in a single undifferentiated loop.
AEGIS~\citep{aegis} introduces a two-agent design where a Searcher retrieves context and a Reproducer generates tests, but relies on multi-agent coordination that risks information loss across boundaries.
Our work occupies a distinct point: we retain a full agent loop \emph{within} each phase while enforcing DPI structure \emph{at phase boundaries}, combining agent adaptability with pipeline focus without multi-agent overhead.

\begin{figure}[t]
    \centering
    \includegraphics[width=\linewidth]{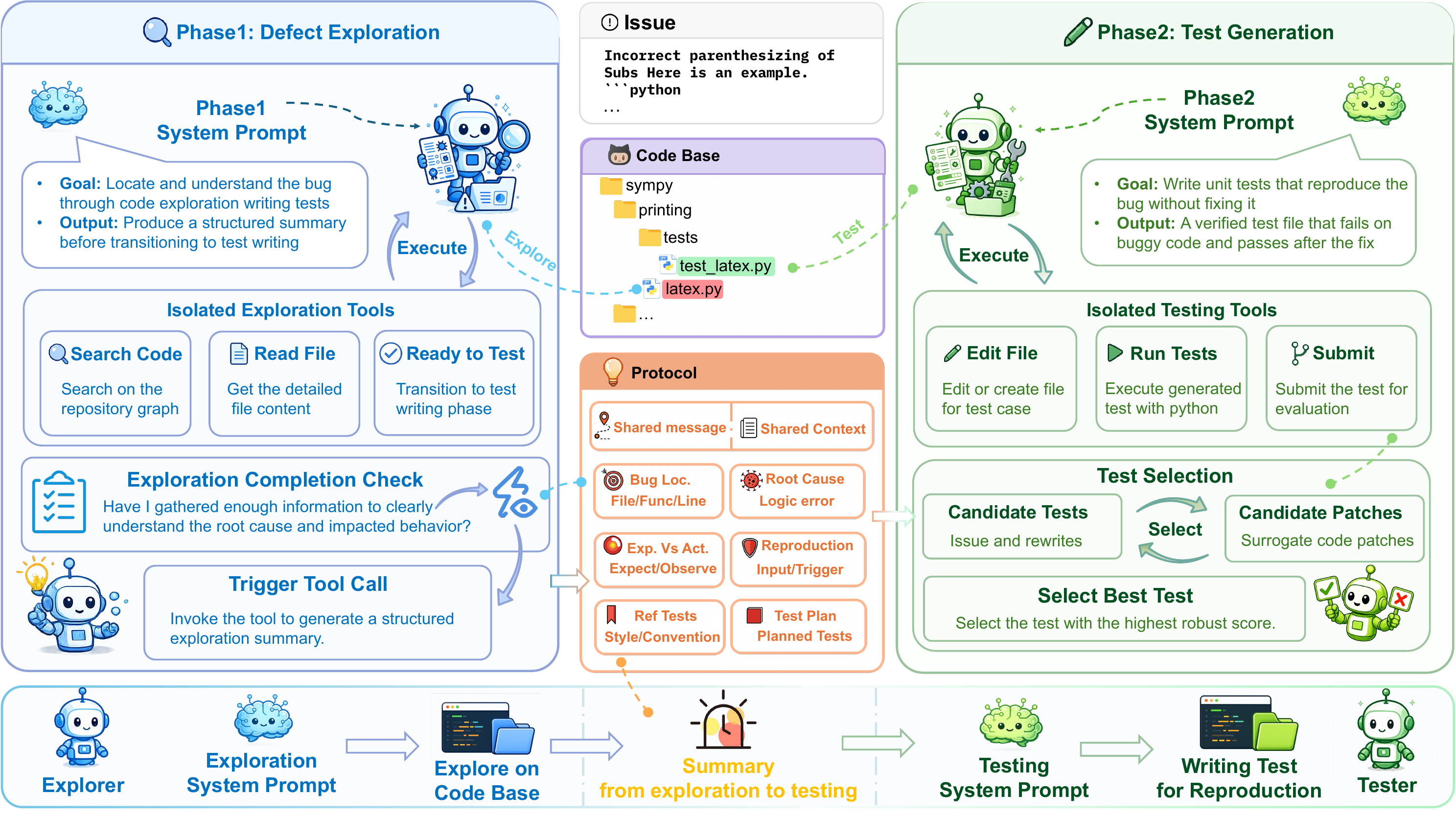}
    \caption{Overview of \mypipeline{}. Following DPI, the pipeline has three stages: \textbf{Divide} splits exploration from test generation; \textbf{Protocol} passes diagnostic findings through a structured summary; and \textbf{Isolate} switches tools across the boundary while preserving context. A final \textbf{Test Selection} stage scores candidates against surrogate patches.}
    \label{fig:pipeline}
    \vspace{-10pt}
\end{figure}

\section{Methodology}
\label{sec:method}
This section details \mypipeline{} as the instantiation of our DPI principles. \S3.1 formulates the problem and overviews the design. Following the \textbf{Divide} principle, \S3.2 and \S3.3 detail the two phases, defect exploration and test generation. \textbf{Protocol} governs the handoff at their boundary, and \textbf{Isolate} constrains each phase's action space to its specific task. \S3.4 introduces the test selection mechanism.

\subsection{Overview}
\paragraph{Problem setup} Given a codebase $C$ and a natural-language-described software issue $I$, the goal of an issue reproduction approach is to automatically generate a test suite $T_{\text{pred}} = \{t_1, t_2, ..., t_M\}$ that reliably reproduces $I$. A generated test $t_i$ is considered to reproduce the issue if it fails on the original buggy codebase $C$ but passes on the fixed codebase $C^\prime$ obtained by applying the golden patch. 
The generated suite $T_{\text{pred}}$ is correct if one $t_i$ reproduces the issue and others show pass-to-pass behavior.

\paragraph{Motivation} While agents generally achieve higher performance than traditional pipelines due to their operational flexibility \citep{yang2025kimidevagentlesstrainingskill}, this uncontrolled freedom becomes a liability in composite tasks like writing reproduction tests. In such unconstrained environments, agents often lose track of their current stage and overall objective, leading to execution confusion and significant fluctuations in the generated results.
To bridge this gap, \mypipeline{} translates the DPI principles (\S\ref{sec:intro}) into a concrete architectural design. By Dividing the monolithic loop, we assign these two subtasks to distinct phases: defect exploration (\S3.2) and test generation (\S3.3). To ensure critical diagnosis context is not lost during the transition, an explicit handoff Protocol bridges the phase boundary. To maintain focus, we Isolate the phases by tailoring the toolset to each phase's specific task, preventing irrelevant tools from misleading the agent's execution. We also use e-Otter++ test selection~\citep{e-otter} (\S3.4) to improve outputs via run variance.

\subsection{Defect Exploration}
In the Defect Exploration phase, the agent commits to building a complete diagnosis of the bug before any test is written. The agent is equipped with file-reading tools such as \texttt{view} and graph-based search tools that operate over a pre-constructed code graph~\cite{rpg-encoder} capturing both code dependencies and semantic relations. The graph enables the agent to retrieve bug-relevant code through structural relationships rather than text patterns, which is faster and more precise.

After exploration, \mypipeline{} enforces the Protocol principle at the phase boundary: the agent must invoke \texttt{ready\_to\_write\_test} to produce a structured handoff summary before transitioning, preventing context loss. The summary captures the agent's diagnosis and a concrete test plan. The diagnosis spans five fields: (1)~bug location, with exact file paths and function names; (2)~root cause, identifying the specific logic error; (3)~expected versus actual behavior; (4)~reproduction strategy, describing the inputs or conditions that trigger the bug; (5)~reference tests, noting existing test files examined for conventions. The test plan (6) outlines which test file to create and what assertions to make. Together, the diagnosis and test plan constitute the agent's complete brief for the test writing phase, ensuring it begins from a finalized understanding rather than partial fragments.

\subsection{Test Generation}
Upon transition, the agent enters the Test Generation phase with the full exploration trajectory preserved in context. The exploration tools are entirely removed from the action space and replaced with \texttt{edit} for modifying test files and \texttt{bash} for executing tests; the system prompt is correspondingly reoriented. This realizes the Isolate principle: with the agent's tools aligned to the test-writing task, irrelevant tools cannot mislead execution, keeping the agent's full Phase 2 effort dedicated to test construction. Combined with the exploration summary as a guideline, the agent focuses on crafting and verifying reproduction tests without repeating prior investigation.

\subsection{Test Selection}
On top of DPI's structural guarantees, individual pipeline runs naturally vary in candidate quality. We adopt a test selection step from e-Otter++~\citep{e-otter} that deliberately diversifies inputs and selects across the resulting candidate pool, turning this variance into improved final output quality.

Following e-Otter++ \citep{e-otter}, we generate five stylistically distinct rewrites of each issue and run \mypipeline on each, yielding a candidate set $\mathcal{T}$ of six test patches. We use SWE-Agent \citep{sweagent} to generate three surrogate patches $\mathcal{B}$, proxying the unavailable golden patch.

Selection proceeds in two steps. First, we \textbf{filter} $\mathcal{T}$ to retain only candidates that exhibit fail-to-pass (F2P) behavior on at least one $b_i \in \mathcal{B}$ (if none qualifies, all candidates are kept as fallback). Second, we \textbf{rank} the surviving candidates by the number of surrogate patches under which F2P holds, with average coverage as a tiebreak:
\begin{equation}
S_j = \big| \{ b_i \in \mathcal{B} : t_j \text{ is F2P under } b_i \} \big|, \qquad \mathrm{cov}_j = \frac{1}{|\mathcal{B}|} \sum_i \mathrm{Coverage}(t_j, C \oplus b_i).
\end{equation}
The top-ranked candidate is returned. The intuition is that a test that triggers the bug under more diverse fixes is more likely to capture the true buggy behavior than to overfit a specific patch.

\begin{table}[!ht]
\centering
\caption{Performance results on SWT-Bench Verified. \textbf{Ours} applies DPI on top of SWE-Agent, \textbf{without test selection}.}
\label{tab:results}
\setlength{\tabcolsep}{8pt}
\renewcommand{\arraystretch}{1.35}
\small
\begin{adjustbox}{max width=\textwidth}
\begin{tabular}{l l R R R R}
\toprule
\textbf{Model} & \textbf{Method} & \textbf{$\mathbf{\mathcal{S}.} \uparrow$} & \textbf{$\mathbf{\Delta C} \uparrow$} & \textbf{$\mathbf{Cov.} \uparrow$} & \textbf{$\mathbf{TDD} \uparrow$} \\
\midrule

\multirow{4}{*}{\colorbox{gpt4o}{\textcolor{gpt4otext}{\texttt{\textbf{GPT-4o}}}}}
& LIBRO$^\dagger$ & 17.80 & 38.00 & -- & -- \\
& AssertFlip$^\dagger$ & 45.50 & 47.40 & -- & -- \\
& Otter$^\dagger$ & 31.60 & 37.60 & 74.59 & 25.78 \\
& Otter++$^\dagger$ & 37.40 & 42.80 & 79.47 & 34.04 \\

\colorbox{claude37}{\textcolor{claude37text}{\texttt{\textbf{Claude-Sonnet-3.7}}}}
& e-Otter++$^\dagger$ & 62.10 & 62.30 & 85.73 & 56.80 \\
\colorbox{amazonq}{\textcolor{amazonqtext}{\texttt{\textbf{v20250405-dev}}}}
& Amazon Q$^\dagger$ & 51.00 &  57.40 & 70.04 & 42.72\\

\colorbox{gemini25p}{\textcolor{gemini25ptext}{\texttt{\textbf{Gemini-2.5-Pro}}}}
& Echo$^\dagger$ & 66.30 & 68.50 & -- & -- \\


\midrule

\multirow{8}{*}{\colorbox{gpt5m}{\textcolor{gpt5mtext}{\texttt{\textbf{GPT-5-Mini}}}}}
& TraeAgent &60.74 & 55.95 & 64.75 & 47.23 \\
& Mini-SWE-Agent &38.80 & 41.54 & 53.11 & 35.02\\
& OpenHands$^\dagger$ & 62.40 & 60.60 & 69.92 & 42.13\\
& Claude Code & 55.65 & 53.11 & 67.62 & 45.69\\
& Aider & 40.42 & 41.33 & 47.62  & 30.62 \\
& Terminus-2 & 42.73 & 50.39 & 60.21 & 35.04 \\
& SWE-Agent & 60.12 & 64.78 & \textcolor{bestval}{\textbf{80.12}} & 54.29\\
& \cellbest{\textbf{Ours}} & \cellbestval{74.28} & \cellbestval{65.51} & \cellbest{75.97} & \cellbestval{62.22}\\

\midrule

\multirow{8}{*}{\colorbox{gpt5}{\textcolor{gpt5text}{\texttt{\textbf{GPT-5}}}}}
& TraeAgent & 69.98 & 61.77 & 75.80 & 60.59\\
& Mini-SWE-Agent & 54.27 & 44.80 & 54.69 & 41.94\\
& OpenHands$^\dagger$ & 79.80 & 66.30 & 68.85 & 45.79\\
& Claude Code & 72.51 & 64.63 & 78.50 & 59.94 \\ 
& Aider & 34.87 & 39.33 & 46.84 &  26.59\\
& Terminus-2 & 44.11 & 48.14 & 58.46 &  34.23\\
& SWE-Agent & 72.52 & 69.26 & 84.47 & 56.30\\
& \cellbest{\textbf{Ours}} & \cellbest{\textcolor{bestval}{\textbf{81.76}}} & \cellbestval{70.95} & \cellbestval{84.92}& \cellbestval{68.02}\\

\midrule

\multirow{8}{*}{\colorbox{claude45}{\textcolor{claude45text}{\texttt{\textbf{Claude-Opus-4.5}}}}}
& TraeAgent & 77.60 & 60.25 & 72.22 & 59.75\\
& Mini-SWE-Agent &76.67& 67.91 & 82.37& 66.85\\
& OpenHands & 73.90 & 57.17 & 71.69& 55.34\\
& Claude Code & 73.21 & 59.48 & 70.70 & 55.35\\
& Aider & 48.26 & 51.29 & 59.59 & 35.97 \\
& Terminus-2 & 68.13 & 54.22 & 59.57 & 41.99 \\
& SWE-Agent & 76.91 & 71.62 & 86.64 & 69.77\\
& \cellbest{\textbf{Ours}} & \cellbestval{78.98} & \cellbestval{71.74} & \cellbestval{89.52}& \cellbestval{73.77}\\

\bottomrule
\end{tabular}
\end{adjustbox}
\par\vspace{2pt}
{\footnotesize\raggedright $^\dagger$Results reported from the public leaderboard.\par}
\vspace{-10pt}
\end{table}

\section{Experiments Setup}
\label{sec:setup}
\paragraph{Benchmark} We evaluate \mypipeline{} on SWT-Bench~\citep{swtbench}, a benchmark constructed from SWE-Bench~\citep{swebench} that organizes GitHub issues together with their corresponding fixes and test cases. 
SWT-Bench filters out instances with unreliable golden patches due to flaky tests and offers two subsets: SWT-Bench Lite (276 instances) and SWT-Bench Verified (433 instances). Consistent with prior work, we use SWT-Bench Verified as our primary evaluation dataset, which provides a larger set of validated instances and reliable comparisons across models.

\paragraph{Baselines} We compare against representative approaches for test generation: SWE-Agent~\citep{sweagent} equips the LLM with direct access to a limited shell environment and provides specialized tools for file searching, viewing, and editing; Mini-SWE-Agent~\citep{sweagent} is a lightweight variant that simplifies the tool interface while maintaining core exploration capabilities; OpenHands~\citep{openhands} is a generalist agent that interacts with the repository via bash commands and Python calls to a predefined skill library within a stateful IPython session; TraeAgent~\citep{traeagent} ensembles multiple candidate patches and selects the final one via pruning and a voting-based selector agent.
Aider\footnote{\url{https://github.com/Aider-AI/aider}} performs repository indexing to guide file selection, incorporates model-generated summaries to augment context, and validates edits via static analysis and repository test cases before application.
Terminus-2~\citep{terminus-2} is a minimal scaffold that interacts with the codebase purely through Bash commands in a headless terminal, designed as a neutral testbed without specialized tools. Claude Code\footnote{\url{https://claude.ai/code}} is Anthropic's official command-line coding agent that operates through a curated set of file and shell tools. Together, these baselines span a diverse range of agent architectures, from minimal terminal scaffolds to specialized SWE agents and ensemble-based systems, providing a comprehensive comparison across the design space of LLM-based test generation.
Implementation details for each baseline are provided in Appendix~\ref{sec:baseline}.

\paragraph{Evaluation Metrics} We adopt three standard metrics following SWT-Bench~\citep{swtbench}: Success rate ($\mathbf{\mathcal{S}}$) measures the fraction of instances for which the generated tests successfully reproduce the target issue; The Mean coverage ($\mathbf{Cov.}$) reports the average fraction of the golden patch's executable modified lines covered by the generated tests. 
The Change coverage ($\mathbf{\Delta C}$) further restricts this measurement to lines that are not already covered by the repository's original test suite, isolating the incremental coverage contributed by the generated tests. 
In addition, we follow~\citep{otter} and include TDD coverage ($\mathbf{TDD}$), which credits coverage only from instances satisfying $\mathcal{S}$, formally
$\mathbf{TDD} = \frac{1}{N} \sum_{i: \mathcal{S}_i=1} Cov._i$,
where $N$ is the total number of instances. This metric discounts tests that touch the modified code without triggering the bug.

\paragraph{Implementation Details} We evaluate \mypipeline{} with three backbone LLMs: GPT-5-Mini~\citep{gpt5-system-card}, GPT-5~\citep{gpt5-system-card}, and Claude-Opus-4.5.
Baselines follow their original configurations with all other parameters kept at default values.
To evaluate the correctness of the generated tests, we adopt the official SWT-Bench Docker environment~\citep{swtbench} to ensure consistent and reliable execution. 
All results are averaged over 3 runs.

\section{Main Results}
\label{main-results}
\paragraph{\mypipeline{} Substantially Improves Issue Reproduction} Table~\ref{tab:results} demonstrates that our method consistently achieves the highest success rate ($\mathcal{S}$) across all three backbone models. On GPT-5-Mini, \mypipeline{} reaches 74.28\% $\mathcal{S}$, outperforming the strongest baseline OpenHands by 11.88 points. Although SWE-Agent reports a higher $Cov.$ of 80.12\%, its $TDD$ drops to 54.29\%, well below our 62.22\%, indicating that much of its coverage comes from tests that do not actually reproduce the issue.
Furthermore, on GPT-5, \mypipeline{} elevates $\mathcal{S}$ to 81.76\%, while simultaneously achieving the highest $Cov.$ and $\Delta C$ among all methods.
Since \mypipeline{} is built on top of SWE-Agent, Figure~\ref{fig:teaser} directly visualizes the gains from DPI: +14.16 $\mathcal{S}$ on GPT-5-Mini, +9.24 on GPT-5, and +2.07 on Claude-Opus-4.5. The $TDD$ improvements (+7.93, +11.72, +4.00) are particularly notable, confirming that DPI helps the agent write tests that genuinely discriminate buggy from fixed behavior rather than merely touching modified lines. The gains are most pronounced on weaker backbones, suggesting that structured decomposition compensates for limited model capability.

\begin{figure}[t]
\centering
\begin{subfigure}[t]{0.32\textwidth}
    \centering
    \includegraphics[width=\linewidth]{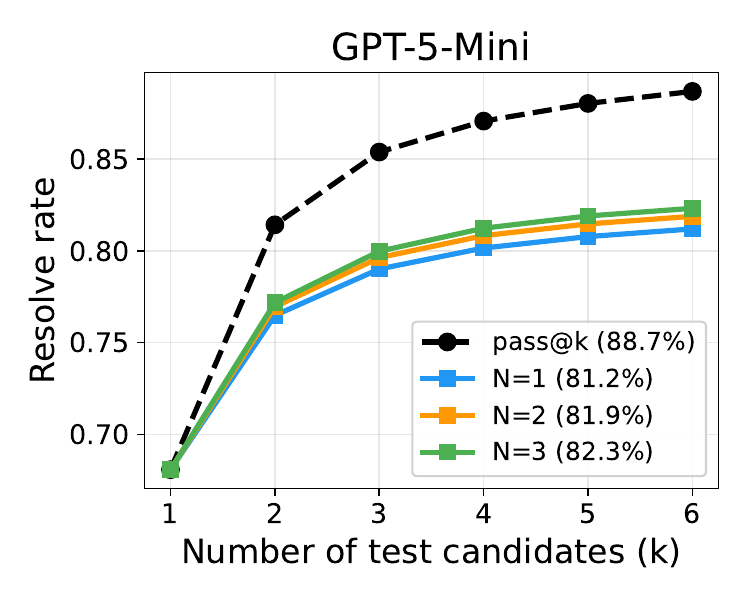}
    \caption{GPT-5-Mini}
    \label{fig:sel-gpt5mini}
\end{subfigure}
\hfill
\begin{subfigure}[t]{0.32\textwidth}
    \centering
    \includegraphics[width=\linewidth]{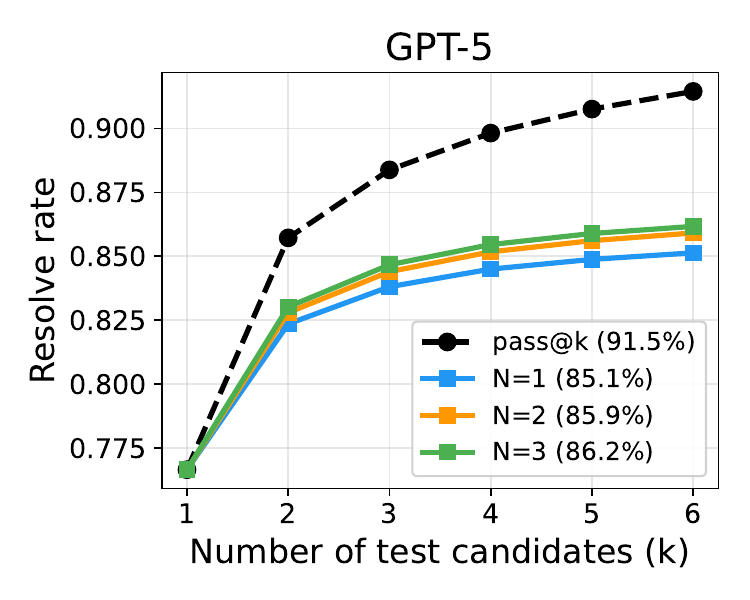}
    \caption{GPT-5}
    \label{fig:sel-gpt5}
\end{subfigure}
\hfill
\begin{subfigure}[t]{0.32\textwidth}
    \centering
    \includegraphics[width=\linewidth]{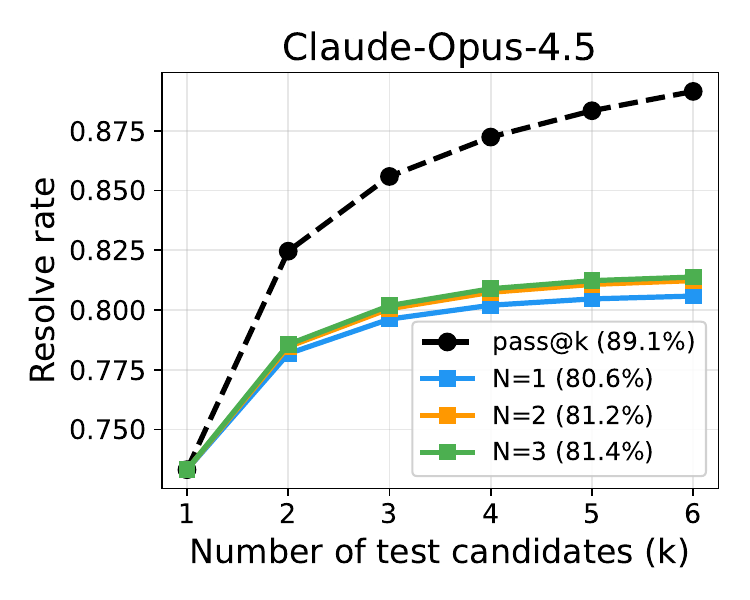}
    \caption{Claude-Opus-4.5}
    \label{fig:sel-claude}
\end{subfigure}
\caption{Resolve rate of test selection using $N$ code patches compared with the pass@k oracle upper bound, as a function of the number of test candidates $k$. $N$ denotes the number of code patches used to score and rank test candidates. The percentages in the legend denote the resolve rate at $k=6$.}
\label{fig:passk-vs-selection}
\vspace{-10pt}
\end{figure}

\paragraph{\revision{DPI Improves Pre-feedback Test Readiness}}
\revision{
To characterize what phase structure contributes before execution feedback, we evaluate each trajectory's first test edit with an LLM-as-Judge.
The judge scores two dimensions:
\emph{Target Alignment} (Target), whether the test targets the correct code location, and
\emph{Specification Completeness} (Spec), whether the test captures the trigger condition, critical input, call path, oracle, and discriminative behavior.
We mark a first edit as \emph{Ready} only when both dimensions receive the maximum score.
We instantiate the judge with GPT-5.6-Terra, run it three times independently, apply outcome-aware correction by marking an edit Ready if it already achieves F2P without post-first-edit modification, and report mean case counts.
Table~\ref{tab:first-edit-quality} shows that DPI primarily improves pre-feedback test design quality, and this improvement is most critical for weaker models.
DPI consistently lifts Ready counts across all three backbones; since Ready cases convert to F2P at over 90\% across all configurations, increasing Ready directly expands the solved set.
Weaker models benefit more because they rarely recover from Not-Ready first edits, whereas stronger models can self-correct more often, so DPI's new Ready cases overlap with cases they could already solve.
}

\begin{table}[!ht]
\centering
\caption{Pre-feedback test design quality on SWT-Bench Verified. Entries are case counts; $\Delta$ rows are count differences.}
\label{tab:first-edit-quality}
\setlength{\tabcolsep}{6pt}
\renewcommand{\arraystretch}{1.25}
\small
\resizebox{\textwidth}{!}{%
\begin{tabular}{l l r r r r r}
\toprule
\textbf{Model} & \textbf{Config} & \textbf{Target} & \textbf{Spec} & \textbf{Ready} & \textbf{Ready F2P} & \textbf{Total F2P} \\
\midrule
\multirow{3}{*}{\colorbox{gpt5m}{\textcolor{gpt5mtext}{\texttt{\textbf{GPT-5-Mini}}}}}
& SWE-Agent & 319.7$\pm$0.6 & 269.7$\pm$1.2 & 269.7$\pm$1.2 & 243.7$\pm$1.5 & 273.7$\pm$1.5 \\
& DPI & 370.0$\pm$1.7 & 332.7$\pm$4.2 & 332.7$\pm$4.2 & 311.0$\pm$2.0 & 331.0$\pm$2.0 \\
& \cellbest{$\Delta$} & \cellbestval{+50.7$\pm$1.2} & \cellbestval{+63.0$\pm$3.5} & \cellbestval{+63.0$\pm$3.5} & \cellbestval{+67.3$\pm$1.5} & \cellbestval{+57.7$\pm$1.2} \\
\midrule
\multirow{3}{*}{\colorbox{gpt5}{\textcolor{gpt5text}{\texttt{\textbf{GPT-5}}}}}
& SWE-Agent & 369.3$\pm$1.5 & 310.0$\pm$4.0 & 310.0$\pm$4.0 & 284.7$\pm$3.5 & 316.7$\pm$2.5 \\
& DPI & 405.0$\pm$1.7 & 377.7$\pm$1.5 & 377.7$\pm$1.5 & 352.7$\pm$0.6 & 355.7$\pm$0.6 \\
& \cellbest{$\Delta$} & \cellbestval{+35.3$\pm$1.2} & \cellbestval{+67.7$\pm$4.7} & \cellbestval{+67.7$\pm$4.7} & \cellbestval{+68.0$\pm$3.6} & \cellbestval{+38.7$\pm$2.9} \\
\midrule
\multirow{3}{*}{\colorbox{claude45}{\textcolor{claude45text}{\texttt{\textbf{Claude-Opus-4.5}}}}}
& SWE-Agent & 365.0$\pm$0.0 & 294.7$\pm$0.6 & 294.7$\pm$0.6 & 269.7$\pm$1.5 & 332.7$\pm$0.6 \\
& DPI & 392.3$\pm$0.6 & 365.0$\pm$2.6 & 365.0$\pm$2.6 & 338.3$\pm$1.2 & 344.3$\pm$1.2 \\
& \cellbest{$\Delta$} & \cellbestval{+27.3$\pm$0.6} & \cellbestval{+70.3$\pm$2.1} & \cellbestval{+70.3$\pm$2.1} & \cellbestval{+68.7$\pm$0.6} & \cellbestval{+11.7$\pm$1.5} \\
\bottomrule
\end{tabular}
}
\vspace{-5pt}
\end{table}

\paragraph{Test Selection Improves Reliability} Figure~\ref{fig:passk-vs-selection} reports the success rate as a function of the number of candidate runs, comparing the oracle upper bound (Pass@$k$) with our surrogate patch-based selection. 
Selection consistently improves as $N$ increases, confirming that additional surrogate patches provide complementary signal for ranking candidates. 
Compared to our pipeline without selection (Table~\ref{tab:results}), adding test selection with $k$=6 and $N$=3 further improves success rate by 8.04 points on GPT-5-Mini (from 74.28\% to 82.32\%), 4.41 points on GPT-5 (from 81.76\% to 86.17\%), and 2.40 points on Claude-Opus-4.5 (from 78.98\% to 81.38\%).
We find that weaker models benefit significantly more from selection than stronger ones. Although Claude-Opus-4.5 has a high Pass@$k$ ceiling (89.15\%), its candidates exhibit subtler behavioral differences that make surrogate patch-based scoring less effective at identifying the best one. We provide a detailed ablation of selection signals in Appendix~\ref{sec:analysis}.

\section{Ablation Study}
\paragraph{Experimental Setup} To isolate the contribution of each design choice, we run component ablations on SWT-Bench Verified with GPT-5-Mini and Claude-Opus-4.5. 
Each ablation removes one or more DPI principles.
For framework ablation (Table~\ref{tab:ablation-framework}), we remove the structured handoff summary (\emph{w/o Prot.}, removing Protocol), expose all tools across phases (\emph{w/o Isol.}, removing Isolate), and collapse the two-stage design into a single reactive loop (\emph{w/o Divd.}, removing Divide) to assess structured planning, phase-specific tool gating, and stage decoupling.

\begin{table}[!ht]
\centering
\caption{Ablation study of DPI principles on SWT-Bench. w/o Prot.\ removes the structured handoff summary, w/o Isol.\ exposes all tools across phases, and w/o Divd.\ collapses the two-stage design into a single reactive loop.}
\label{tab:ablation-framework}
\renewcommand{\arraystretch}{1.25}
\setlength{\tabcolsep}{6pt}
\resizebox{\textwidth}{!}{%
\begin{tabular}{@{}l cccc cccc@{}}
\toprule
\multirow{2}{*}{\textbf{Metric}}
& \multicolumn{4}{c}{\colorbox{gpt5m}{\textcolor{gpt5mtext}{\texttt{\textbf{GPT-5-Mini}}}}}
& \multicolumn{4}{c}{\colorbox{claude45}{\textcolor{claude45text}{\texttt{\textbf{Claude-Opus-4.5}}}}} \\
\cmidrule(lr){2-5}\cmidrule(lr){6-9}
& \textbf{Full} & \textbf{w/o Prot.} & \textbf{w/o Isol.} & \textbf{w/o Divd.} & \textbf{Full} & \textbf{w/o Prot.} & \textbf{w/o Isol.} & \textbf{w/o Divd.} \\
\midrule
$\mathbf{\mathcal{S}.} \uparrow$ & \cellbestval{\textbf{74.28}} & 70.90 & 72.70 & 67.66 & \cellbestval{\textbf{78.98}} & 75.51 & 76.21 & 69.05 \\
$\mathbf{\Delta C} \uparrow$     & \cellbestval{\textbf{65.51}} & 63.71 & 59.40 & 54.59 & \cellbestval{\textbf{71.74}} & 65.98 & 64.03 & 61.03 \\
$\mathbf{Cov.} \uparrow$         & \cellbestval{\textbf{75.97}} & 72.88 & 67.80 & 64.28 & \cellbestval{\textbf{89.52}} & 79.27 & 79.50 & 75.96 \\
\bottomrule
\end{tabular}%
}
\vspace{-5pt}
\end{table}

\paragraph{DPI Components Play Complementary Roles} Table~\ref{tab:ablation-framework} reveals the distinct contributions of each framework component. 
The Protocol component (structured handoff summary) transforms the agent's multi-turn exploration into a detailed plan that guides downstream test generation. 
Removing it consistently degrades performance, with $\mathcal{S}$ dropping from 78.98\% to 75.51\% on Claude-Opus-4.5.
Isolate primarily affects test quality: exposing all tools across phases reduces $\Delta C$ by 6.11 points and $Cov.$ by 8.17 points on GPT-5-Mini, and by 7.71 and 10.02 points on Claude-Opus-4.5.
Divide separates repository understanding from test generation, allowing each phase to focus on a single objective. Collapsing it into a single reactive loop causes substantial drops across all metrics, with $\mathcal{S}$ falling by 9.93 points and $Cov.$ dropping by 13.56 points on Claude-Opus-4.5.
The full pipeline integrates all three principles to consistently outperform all ablated variants across both backbones.

\begin{figure}[h]
\begin{minipage}{0.46\columnwidth}
    \centering
    \includegraphics[width=\linewidth]{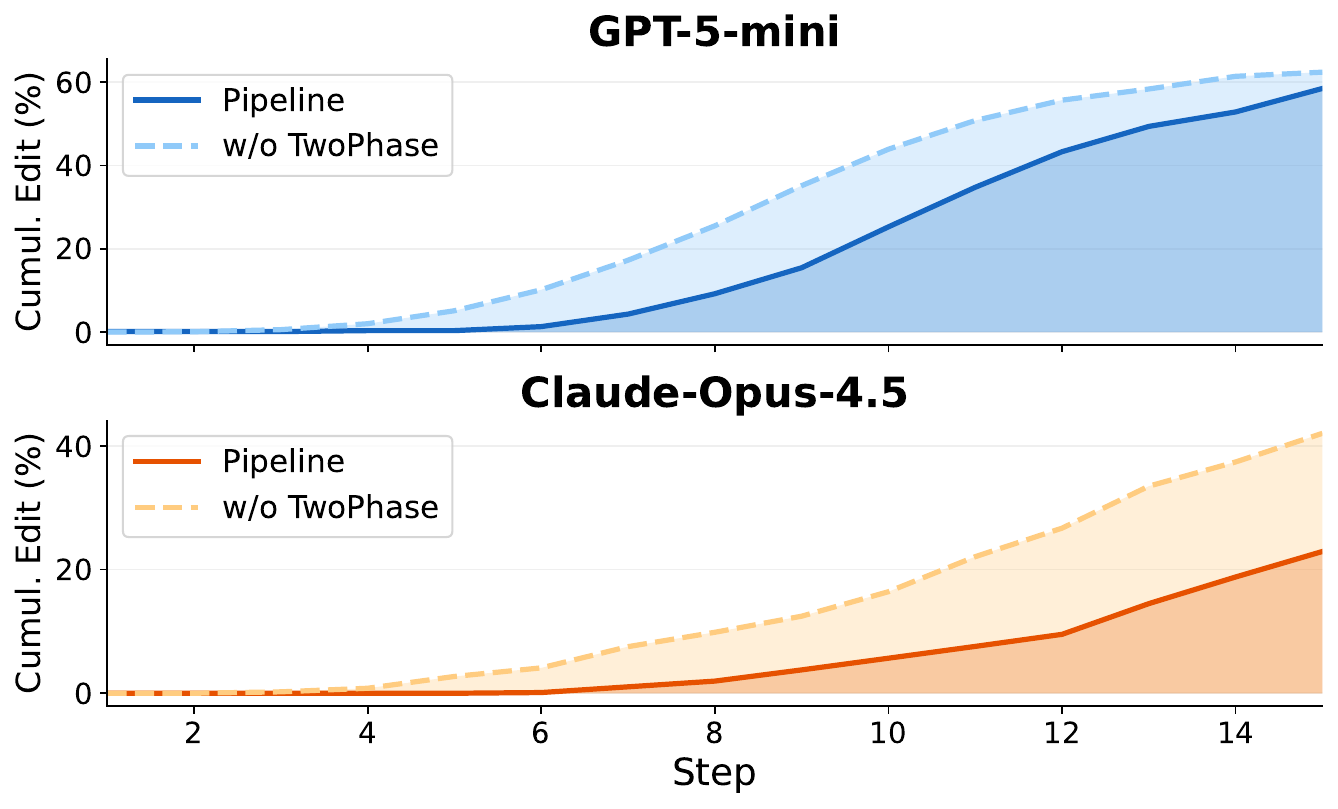}
    \captionof{figure}{Edit tool usage rate across steps for Pipeline and w/o Divd.\ variant.}
    \label{fig:edit-tool}
\end{minipage}
\hfill
\begin{minipage}{0.50\columnwidth}
    \centering
    \renewcommand{\arraystretch}{1.2}
    \setlength{\tabcolsep}{4pt}
    \captionof{table}{Attribution study on GPT-5-Mini. Tool Contribution rows add individual mechanisms to monolithic SWE-Agent; + DPI applies DPI phase structure with standard SWE-Agent tools and no Search Code.}
    \label{tab:phase-attribution}
    \resizebox{\linewidth}{!}{%
    \begin{tabular}{@{}llrrrr@{}}
    \toprule
    \textbf{Type} & \textbf{Config.} & $\mathbf{\mathcal{S}.} \uparrow$ & $\mathbf{TDD} \uparrow$ & $\Delta\mathbf{\mathcal{S}.}$ & $\Delta\mathbf{TDD}$ \\
    \midrule
    Baseline & SWE-Agent & 60.12 & 54.29 & -- & -- \\
    Tool Contribution & + Search Code & 67.66 & 50.40 & +7.54 & -3.89 \\
     & + Summary Prompt & 54.96 & 42.44 & -5.16 & -11.85 \\
     & + Transition Tool & 61.20 & 48.68 & +1.08 & -5.61 \\
    \cellbest{DPI Contribution} & \cellbest{+ DPI} & \cellbestval{\textbf{72.97}} & \cellbestval{\textbf{56.41}} & \cellbestval{\textbf{+12.85}} & \cellbestval{\textbf{+2.12}} \\
    \bottomrule
    \end{tabular}%
    }
\end{minipage}
\end{figure}

\paragraph{\revision{DPI Gains Come from Phase Structure Rather Than Tool Changes}}
\revision{
To disentangle architectural structure from implementation differences, we add individual mechanisms from DPI to the monolithic SWE-Agent baseline on GPT-5-Mini (Table~\ref{tab:phase-attribution}).
Figure~\ref{fig:edit-tool} shows that without the two-phase design, the agent begins invoking edit tools from the very first steps, interleaving test writing with repository exploration rather than first building a complete understanding of the codebase.
Adding the graph-based \texttt{Search Code} tool improves success rate but lowers TDD coverage, indicating that better retrieval alone does not produce more discriminative reproduction tests.
Similarly, injecting the six-field structured summary as a one-shot prompt or exposing \texttt{ready\_to\_write\_test} as a standalone tool does not reproduce DPI's gain.
In contrast, applying the DPI phase structure with only standard SWE-Agent tools achieves substantially larger improvement, showing that the executable phase boundary is the primary source of the gain.
}

\section{Analysis}
Beyond ablation, we analyze agent efficiency from two angles: the effect of \textbf{exploration tools} on step count and \textbf{convergence speed} relative to baselines. Additional repository-level and tool-usage analyses are provided in Appendix~\ref{sec:analysis}.

\begin{figure}[h]
\begin{minipage}{0.46\columnwidth}
    \centering
    \includegraphics[width=\linewidth]{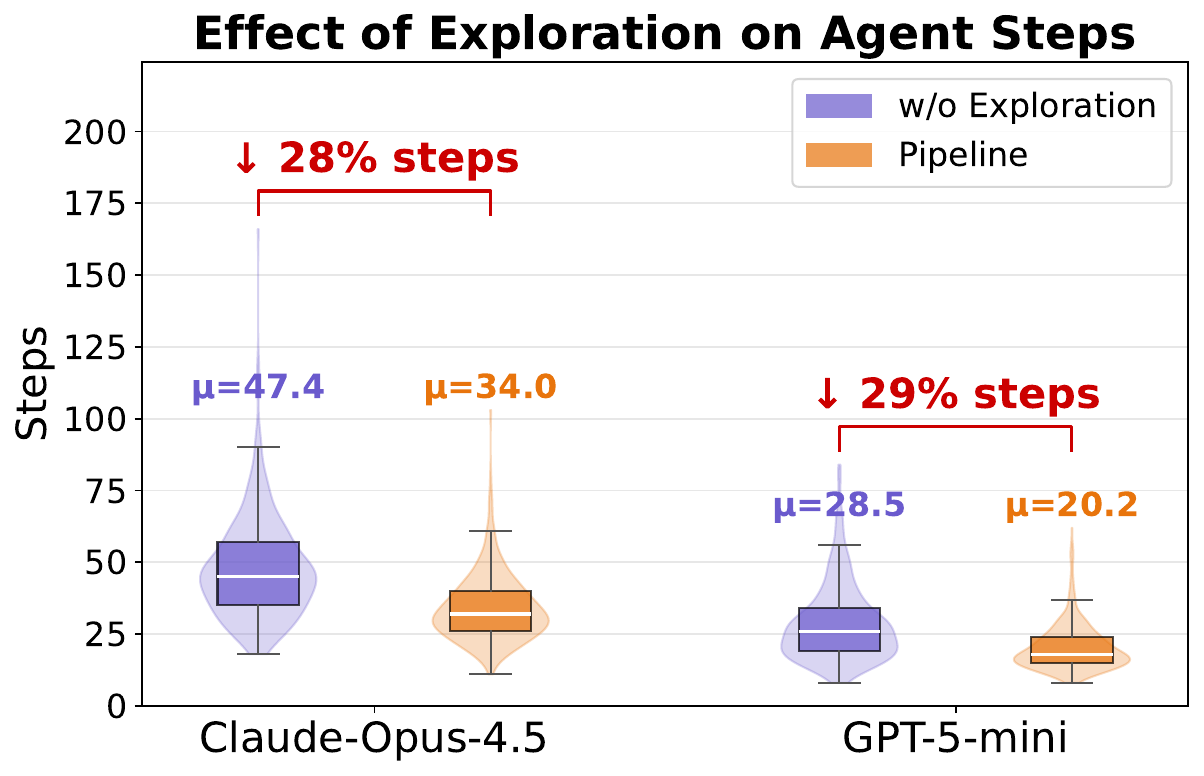}
    \captionof{figure}{Distribution of agent steps with and without exploration tools.}
    \label{fig:steps-opus-5mini}
\end{minipage}
\hfill
\begin{minipage}{0.50\columnwidth}
    \centering
    \renewcommand{\arraystretch}{1.3}
    \setlength{\tabcolsep}{5pt}
    \captionof{table}{Tool ablation results. w/o Expl. denotes the variant without efficient exploration tools.}
    \label{tab:ablation-rpg}
    \resizebox{\linewidth}{!}{%
    \begin{tabular}{@{}l cc cc@{}}
    \toprule
    \multirow{2}{*}{\textbf{Metric}}
    & \multicolumn{2}{c}{\colorbox{gpt5m}{\textcolor{gpt5mtext}{\texttt{\textbf{GPT-5-Mini}}}}}
    & \multicolumn{2}{c}{\colorbox{claude45}{\textcolor{claude45text}{\texttt{\textbf{Claude-Opus-4.5}}}}} \\
    \cmidrule(lr){2-3}\cmidrule(lr){4-5}
    & \textbf{Full} & \textbf{w/o Expl.} & \textbf{Full} & \textbf{w/o Expl.} \\
    \midrule
    $\mathbf{\mathcal{S}.} \uparrow$ & \cellbestval{\textbf{74.28}} & 72.97 & \cellbestval{\textbf{78.98}} & 73.67 \\
    $\mathbf{\Delta C} \uparrow$     & \cellbestval{\textbf{65.51}} & 62.26 & \cellbestval{\textbf{71.74}} & 67.10 \\
    $\mathbf{Cov.} \uparrow$         & \cellbestval{\textbf{75.97}} & 71.18 & \cellbestval{\textbf{89.52}} & 80.98 \\
    \bottomrule
    \end{tabular}%
    }
\end{minipage}
\end{figure}

\paragraph{Efficient Exploration Tools Enable Better Repository Understanding} Table~\ref{tab:ablation-rpg} examines the contribution of our exploration tools, which provide structured access to the repository through graph navigation. 
Replacing them with \texttt{str\_replace\_editor} and \texttt{bash} as in SWE-Agent leads to consistent drops across all metrics: on Claude-Opus-4.5, $\mathcal{S}$ falls from 78.98\% to 73.67\% and $Cov.$ drops by 8.54 points. 
These results indicate that structured repository access through graph navigation significantly improves test generation performance across all metrics.
Beyond test generation performance, as shown in Figure~\ref{fig:steps-opus-5mini}, exploration tools also improve agent efficiency by reducing total steps by 29\% across all backbones. Please refer to Appendix~\ref{sec:analysis} for details.

\begin{figure}[!ht]
\centering
\begin{subfigure}[t]{0.32\textwidth}
    \centering
    \includegraphics[width=\linewidth]{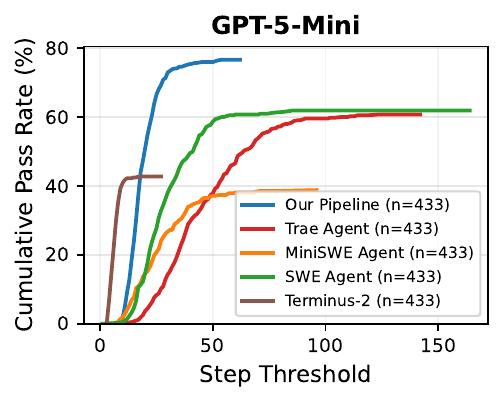}
    \caption{GPT-5-Mini}
    \label{fig:pass-gpt5mini}
\end{subfigure}
\hfill
\begin{subfigure}[t]{0.32\textwidth}
    \centering
    \includegraphics[width=\linewidth]{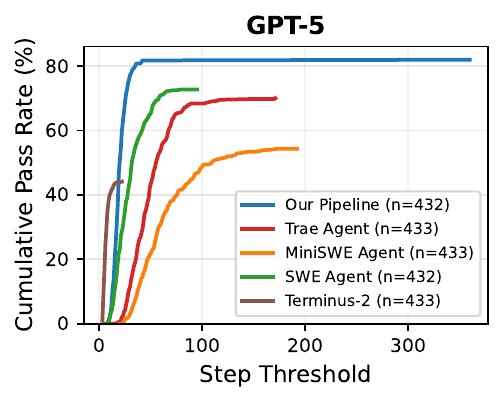}
    \caption{GPT-5}
    \label{fig:pass-gpt5}
\end{subfigure}
\hfill
\begin{subfigure}[t]{0.32\textwidth}
    \centering
    \includegraphics[width=\linewidth]{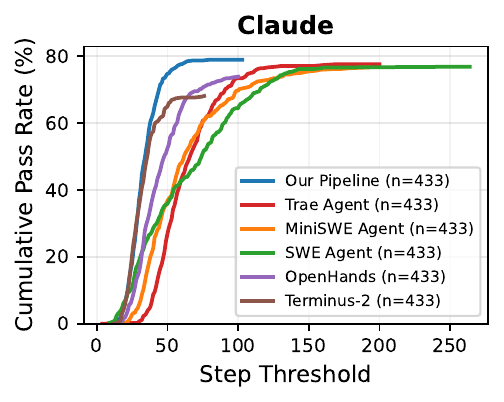}
    \caption{Claude-Opus-4.5}
    \label{fig:pass-claude}
\end{subfigure}
\caption{Cumulative pass rate as a function of the step budget across three backbones. }
\label{fig:pass-vs-steps}
\vspace{-10pt}
\end{figure}

\paragraph{Structured Exploration Boosts Pass-Rate Efficiency} \mypipeline{} converges to higher pass rates with significantly fewer steps than all baselines (Figure~\ref{fig:pass-vs-steps}).  On GPT-5, our pipeline reaches nearly 80\% pass rate within 50 steps, while Trae Agent and MiniSWE Agent take over 150 steps and still converge much lower.  A similar pattern holds on GPT-5-Mini: our pipeline exceeds 70\% pass rate within 60 steps, surpassing baselines that need 2$\times$ more steps yet converge lower.

\section{Conclusion}
In this work, we introduce \mypipeline{}, an agent design for reproduction test generation built on three principles: Divide, Protocol, Isolate (DPI). 
Divide separates exploration from generation into two phases with distinct objectives. Protocol forces the agent to produce a structured diagnosis and test plan before transitioning. Isolate removes wrong-phase tools so phase boundaries are enforced at the action-space level. 
Our evaluations show that DPI consistently improves reproduction success across diverse backbones and baselines. Ultimately, \mypipeline{} establishes that minimal structural constraints at phase boundaries can yield greater gains than scaling backbone capability alone, offering a design blueprint for structured agentic reasoning in software engineering.


\bibliographystyle{unsrtnat}
\bibliography{refs}

\clearpage
\appendix

\section{Detailed Methodology of \mypipeline{}}
\label{sec:detail-method}
This section provides a deep dive into the implementation details of \mypipeline{}, expanding upon the three core components—Exploration, Summary, and Test Generation—introduced in Section~\ref{sec:method} of the main text.

\begin{figure}[!h]
    \centering
    \includegraphics[width=\columnwidth]{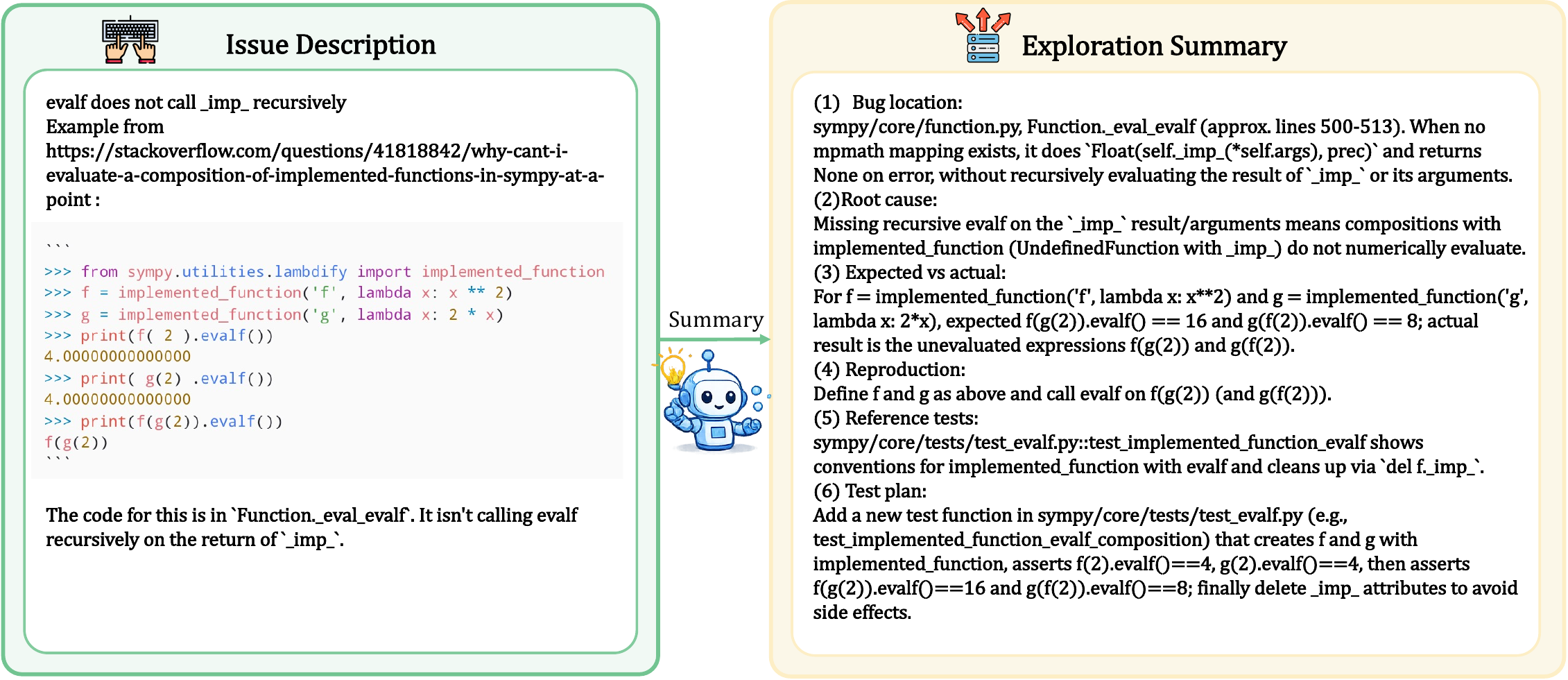}
    \caption{Case study on the issue \texttt{sympy/sympy-12096}: the generated exploration summary demonstrates how defect-relevant information is structured from the raw issue description.}
    \label{fig:summary-case}
\end{figure}

\paragraph{Exploration Phase} The exploration phase is responsible for building a thorough understanding of the codebase relevant to the target issue before any test is written. The agent operates with the system prompt $\textsc{SysPrompt}_{\text{exp}}$ and the exploration tool set $\mathcal{T}_{\text{exp}}$, which combines graph-based search tools (e.g., \texttt{search\_code}) backed by a pre-built repository graph with standard file-viewing and shell utilities (\texttt{str\_replace\_editor view}, \texttt{bash}). The graph-based tools allow the agent to traverse function-level dependencies and locate semantically related code without resorting to text-based searches across the entire codebase. Throughout this phase, the agent iteratively queries the repository, inspects relevant files, and accumulates findings in its trajectory $h$. The phase terminates when the agent decides it has gathered sufficient context and invokes \texttt{ready\_to\_write\_test} with a structured summary, triggering the transition to the test generation phase. The summary consolidates the agent's findings into six fields: \emph{bug location}, \emph{root cause}, \emph{expected vs.\ actual behavior}, \emph{reproduction}, \emph{reference tests}, and \emph{test plan}. Figure~\ref{fig:summary-case} illustrates a concrete example produced for a real issue from \texttt{sympy.geometry}, where the agent localizes the bug in \texttt{Point.distance} to a \texttt{zip} call that silently truncates mismatched-dimension inputs and proposes a test plan with concrete assertions against the buggy behavior.

\paragraph{System Prompt Template for Exploration Phase} The system prompt $\textsc{SysPrompt}_{\text{exp}}$ used during the exploration phase is shown below. It instructs the agent to leverage the graph-based tools for systematic codebase exploration and to invoke \texttt{ready\_to\_write\_test} once it has gathered enough context to produce the structured summary.

\begin{tcblisting}{
  title={Exploration Phase System Prompt ($\textsc{SysPrompt}_{\text{exp}}$)},
  colback=lightgray,
  colframe=black,
  arc=1mm,
  boxrule=1pt,
  left=1mm,right=1mm,top=1mm,bottom=1mm,
  breakable,
  fontupper=\scriptsize\ttfamily,
  listing only,
  listing engine=listings,
  listing options={
    breaklines=true,
    breakatwhitespace=false,
    breakindent=0pt,
    prebreak=\mbox{},
    postbreak=\mbox{},
    keepspaces=true,
    columns=fullflexible,
    tabsize=4
  }
}
## Role
You are a senior software engineer specializing in writing high-quality unit tests. You will receive a bug report plus tools to inspect and edit the source repository.

## Task
Your goal is to write unit tests that **reproduce the described issue**. The tests you write should:
- **Fail** in the current (buggy) state of the repository
- **Pass** once the issue has been resolved

You are NOT expected to fix the bug -- only to write tests that demonstrate it.

## Workflow
Your work has two phases:
1. **Exploration Phase** (current): Use RPG graph tools and bash to understand the bug, locate the relevant code, and reproduce the issue. When you have a clear understanding, call `ready_to_write_test` to transition.
2. **Test Writing Phase**: Write and verify your tests using str_replace_editor and bash.

You MUST call `ready_to_write_test` before you start creating test files.

## Repository Unified Graph (RUG)
You can navigate two connected graphs:
1) Functionality SubGraph ("What"): Domain -> Module -> Functional Behavior
2) Dependency SubGraph ("How"): Imports/Calls/Inherits/Routes/Entry points

## Mandatory Thinking Protocol
Every tool call has a `thinking` parameter. This is your ONLY opportunity to reason explicitly -- use it for deep, structured analysis, NOT shallow one-line summaries.
Each `thinking` MUST contain ALL four sections below. Omitting any section is a failure:
1. **PROGRESS**: Summarize what you have learned from ALL previous steps. List entities found, hypotheses confirmed/rejected, and dead ends. If this is your first step, analyze the bug report anchors instead.
2. **CURRENT HYPOTHESIS**: State your current theory about the bug's root cause and how to reproduce it in a test. Identify what phase you are in (understanding the bug, locating relevant code, reproducing the bug, ready to write tests).
3. **TOOL CHOICE JUSTIFICATION**: Explain why the tool you chose is the best fit for your current need -- and why the other tools are NOT appropriate right now.
4. **EXPECTED OUTCOME & NEXT STEP**: What information do you expect? What will you do next depending on the result? Consider both success and failure scenarios.

## Action Space -- Exploration Phase Tools

## Tool Name: search_code
### Description
- Search and retrieve concrete code from the repository using file paths, qualified names (file:Class.method), or raw text keywords.
- Supports direct symbol lookup and keyword-based code search, and can return full files, specific functions, or targeted lines.
- **LIMITATION**: This tool CANNOT access test files or test directories (e.g., `tests/`, `test/`, `*_test.py`, `test_*.py`). To view test files, use `str_replace_editor` or `bash` instead.
### Parameters
{
    "tool_name": "search_code",
    "parameters": {
        "thinking": "<Your structured reasoning: PROGRESS, CURRENT HYPOTHESIS, TOOL CHOICE JUSTIFICATION, EXPECTED OUTCOME & NEXT STEP>",
        "search_terms": "<List of file paths, qualified code entities, or text keywords>",
        "line_nums": "<List of two integers [start, end] to extract lines from a specific file. Optional.>",
        "file_pattern": "<File path or glob pattern to restrict search scope. Default: '**/*.py'>"
    }
}
### Returns
Matched code snippets, complete files, or located entities based on search terms or line numbers.

## Tool Name: str_replace_editor
### Description
Custom editing tool for viewing, creating, and editing files.
* **IMPORTANT**: In the Exploration Phase, this tool is for VIEWING files only. Do NOT use `create`, `str_replace`, or `insert` to write test files. You MUST call `ready_to_write_test` first to transition to the Test Writing Phase before creating or editing any test files. If you skip this step, you will NOT have access to `submit` and cannot complete the task.
* If `path` is a file, `view` displays the result of applying `cat -n`. If `path` is a directory, `view` lists non-hidden files up to 2 levels deep.
* The `create` command cannot be used if the specified `path` already exists as a file.
* `old_str` must match EXACTLY one or more consecutive lines from the original file.
### Parameters
{
    "tool_name": "str_replace_editor",
    "parameters": {
        "thinking": "<Your structured reasoning: PROGRESS, CURRENT HYPOTHESIS, TOOL CHOICE JUSTIFICATION, EXPECTED OUTCOME & NEXT STEP>",
        "command": "<One of: 'view', 'create', 'str_replace', 'insert', 'undo_edit'>",
        "path": "<Absolute path to file or directory>",
        "file_text": "<Content for 'create' command. Optional.>",
        "old_str": "<String to replace. Optional.>",
        "new_str": "<Replacement string. Optional.>",
        "insert_line": "<Line number after which to insert. Optional.>",
        "view_range": "<Two integers [start, end] for 'view' command. Optional.>"
    }
}

## Tool Name: bash
### Description
Run a shell command in the repository environment. Use this to run tests, install dependencies, explore the filesystem, or execute Python scripts.
* You cannot use interactive commands.
### Parameters
{
    "tool_name": "bash",
    "parameters": {
        "thinking": "<Your structured reasoning: PROGRESS, CURRENT HYPOTHESIS, TOOL CHOICE JUSTIFICATION, EXPECTED OUTCOME & NEXT STEP>",
        "command": "<The shell command to execute>"
    }
}

## Tool Name: ready_to_write_test
### Description
Signal that you have finished exploring and are ready to write tests.
You MUST call this tool before creating or editing test files. After calling this tool, your available tools will switch to test-writing mode (RPG tools will no longer be available).
Your summary MUST be detailed and thorough -- it is the bridge between your exploration and test writing. Include ALL of the following:
1. **Bug location**: Which file(s) and function(s) contain the bug (exact paths and line ranges).
2. **Root cause**: What exactly is wrong in the code (the specific logic error, missing check, wrong return value, etc.).
3. **Expected vs actual behavior**: What should happen vs what currently happens.
4. **Reproduction**: How to trigger the bug (specific inputs, conditions, or API calls).
5. **Reference tests**: Which existing test file(s) and test function(s) you examined for conventions (file path, import patterns, fixture usage, assertion style).
6. **Test plan**: What test file to create (path), what test functions to write, what assertions to make, following the conventions from the reference tests.
### Parameters
{
    "tool_name": "ready_to_write_test",
    "parameters": {
        "thinking": "<Your structured reasoning: PROGRESS, CURRENT HYPOTHESIS, TOOL CHOICE JUSTIFICATION, EXPECTED OUTCOME & NEXT STEP>",
        "summary": "<Detailed summary covering: bug location, root cause, expected vs actual behavior, reproduction steps, reference tests you examined, and test plan. Must be 200-2000 chars.>"
    }
}

## Tool Name: python_add_description
### Description
Add a description to a previously executed Python code record. After executing Python code via bash, you will see a NOTE with an execution ID. Use this tool to annotate that execution.
### Parameters
{
    "tool_name": "python_add_description",
    "parameters": {
        "thinking": "<Your structured reasoning: PROGRESS, CURRENT HYPOTHESIS, TOOL CHOICE JUSTIFICATION, EXPECTED OUTCOME & NEXT STEP>",
        "execution_id": "<The execution ID shown in the NOTE>",
        "description": "<Description of why you ran this code and what the results mean>"
    }
}

## Output Format
Your every response must contain exactly one "<action>...</action>" block:
<action>
{
  "tool_name": "...",
  "parameters": {
    "thinking": "your reasoning here",
    ...
  }
}
</action>
\end{tcblisting}

\paragraph{Instance Template for Exploration Phase} 
The instance template injects the repository path \texttt{\{\{working\_dir\}\}} and the issue description \texttt{\{\{problem\_statement\}\}} into each task instance. It guides the agent to first develop an understanding of the target issue through exploration, including locating relevant code, examining existing test conventions, and reproducing the issue, before transitioning to test writing via \texttt{ready\_to\_write\_test}.

\begin{tcblisting}{
  title={Exploration Phase Instance Template},
  colback=lightgray,
  colframe=black,
  arc=1mm,
  boxrule=1pt,
  left=1mm,right=1mm,top=1mm,bottom=1mm,
  breakable,
  fontupper=\scriptsize\ttfamily,
  listing only,
  listing engine=listings,
  listing options={
    breaklines=true,
    breakatwhitespace=false,
    breakindent=0pt,
    prebreak=\mbox{},
    postbreak=\mbox{},
    keepspaces=true,
    columns=fullflexible,
    tabsize=4
  }
}
<uploaded_files>
{{working_dir}}
</uploaded_files>
I've uploaded a python code repository in the directory {{working_dir}}. We have received the following issue within our repository. Here's the issue text:

<pr_description>
{{problem_statement}}
</pr_description>

INSTRUCTIONS:
Your task is to **write unit tests that reproduce the described issue**. The tests should:
- **Fail** in the current (buggy) state of the repository
- **Pass** once the issue has been resolved

You are NOT expected to fix the bug -- only write tests that demonstrate it.

IMPORTANT TIPS:
1. Start by understanding the bug: read the issue carefully, locate the relevant code, and understand the expected vs. actual behavior.
2. If the issue includes reproduction code, try running it first to confirm the bug exists.
3. Look at existing tests in the repository to follow the same patterns and conventions.
4. When you have a clear understanding of the bug and know what to test, call `ready_to_write_test` to switch to test-writing mode.
5. Do NOT modify any non-test files. Only create or edit test files.
\end{tcblisting}

\paragraph{Test Case Generation Phase}  Once the transition is triggered, the agent enters the test generation phase with a switched configuration: the system prompt is replaced with $\textsc{SysPrompt}_{\text{gen}}$ and the tool set is replaced with $\mathcal{T}_{\text{gen}}$, which retains \texttt{str\_replace\_editor} and \texttt{bash} but removes the graph-based search tools. The exploration trajectory and the structured summary remain in the context $h$, providing the agent with a complete view of the codebase as well as a concrete plan to execute. Guided by the test plan in the summary, the agent creates the target test file at the prescribed path, writes test functions following the conventions identified in the reference tests, and uses \texttt{python} to validate that the generated tests behave as expected against the buggy codebase. The phase terminates when the agent invokes \texttt{submit} with the final test patch $P$.

\paragraph{System Prompt Template for Test Generation Phase} The system prompt $\textsc{SysPrompt}_{\text{gen}}$ replaces $\textsc{SysPrompt}_{\text{exp}}$ after the agent calls \texttt{ready\_to\_write\_test}. It removes all graph tools and provides the agent with file editing, bash execution. The agent writes tests following the plan from its structured summary, verifies them against the buggy code, and calls \texttt{submit} to finalize the patch.

\begin{tcblisting}{
  title={Test Generation Phase System Prompt ($\textsc{SysPrompt}_{\text{gen}}$)},
  colback=lightgray,
  colframe=black,
  arc=1mm,
  boxrule=1pt,
  left=1mm,right=1mm,top=1mm,bottom=1mm,
  breakable,
  fontupper=\scriptsize\ttfamily,
  listing only,
  listing engine=listings,
  listing options={
    breaklines=true,
    breakatwhitespace=false,
    breakindent=0pt,
    prebreak=\mbox{},
    postbreak=\mbox{},
    keepspaces=true,
    columns=fullflexible,
    tabsize=4
  }
}
## Role
You are a senior software engineer writing unit tests to reproduce a reported bug.

## Task
Write unit tests that **fail** in the current (buggy) state and **pass** once the issue is resolved. You are NOT expected to fix the bug.

## Mandatory Thinking Protocol
Every tool call has a `thinking` parameter. Each `thinking` MUST contain:
1. **PROGRESS**: What you have learned and done so far.
2. **CURRENT HYPOTHESIS**: Your theory about the bug and what test will reproduce it.
3. **TOOL CHOICE JUSTIFICATION**: Why this tool is the best choice now.
4. **EXPECTED OUTCOME & NEXT STEP**: What you expect and what comes next.

## Action Space -- Test Writing Tools

## Tool Name: str_replace_editor
### Description
Custom editing tool for viewing, creating, and editing files.
* If `path` is a file, `view` displays the result of applying `cat -n`. If `path` is a directory, `view` lists non-hidden files up to 2 levels deep.
* The `create` command cannot be used if the specified `path` already exists as a file.
* `old_str` must match EXACTLY one or more consecutive lines from the original file.
### Parameters
{
    "tool_name": "str_replace_editor",
    "parameters": {
        "thinking": "<Your structured reasoning: PROGRESS, CURRENT HYPOTHESIS, TOOL CHOICE JUSTIFICATION, EXPECTED OUTCOME & NEXT STEP>",
        "command": "<One of: 'view', 'create', 'str_replace', 'insert', 'undo_edit'>",
        "path": "<Absolute path to file or directory>",
        "file_text": "<Content for 'create' command. Optional.>",
        "old_str": "<String to replace. Optional.>",
        "new_str": "<Replacement string. Optional.>",
        "insert_line": "<Line number after which to insert. Optional.>",
        "view_range": "<Two integers [start, end] for 'view' command. Optional.>"
    }
}

## Tool Name: bash
### Description
Run a shell command in the repository environment.
### Parameters
{
    "tool_name": "bash",
    "parameters": {
        "thinking": "<Your structured reasoning: PROGRESS, CURRENT HYPOTHESIS, TOOL CHOICE JUSTIFICATION, EXPECTED OUTCOME & NEXT STEP>",
        "command": "<The shell command to execute>"
    }
}


## Tool Name: python_add_description
### Description
Add a description to a previously executed Python code record.
### Parameters
{
    "tool_name": "python_add_description",
    "parameters": {
        "thinking": "<Your structured reasoning: PROGRESS, CURRENT HYPOTHESIS, TOOL CHOICE JUSTIFICATION, EXPECTED OUTCOME & NEXT STEP>",
        "execution_id": "<The execution ID>",
        "description": "<Description of the execution>"
    }
}

## Tool Name: submit
### Description
Submit your changes when you are satisfied with the test(s) you have written. This ends the task.
### Parameters
{
    "tool_name": "submit",
    "parameters": {
        "thinking": "<Your structured reasoning: PROGRESS, CURRENT HYPOTHESIS, TOOL CHOICE JUSTIFICATION, EXPECTED OUTCOME>"
    }
}

## Output Format
Your every response must contain exactly one "<action>...</action>" block:
<action>
{
  "tool_name": "...",
  "parameters": {
    "thinking": "your reasoning here",
    ...
  }
}
</action>

IMPORTANT TIPS:
1. Write focused tests that clearly demonstrate the issue -- avoid overly broad or unrelated assertions.
2. Run your tests to verify they fail as expected. Note: `pytest` is often NOT installed in the environment -- check what test runner is available.
3. When you're satisfied, run tests one final time, then use `submit` to finish.
4. Do NOT modify any non-test files. Only create or edit test files.
\end{tcblisting}

Together, these two phases form a clean separation between repository understanding and test writing, with the structured summary serving as the bridge that carries exploration findings into the generation stage. The full procedure is summarized in Algorithm~\ref{alg:pipeline}.

\begin{algorithm}[h]
\caption{The pseudo-code of test patch generation in \mypipeline{}}
\label{alg:pipeline}
\begin{algorithmic}[1]
\Require Codebase $C$, issue $I$, exploration tools $\mathcal{T}_{\text{exp}}$, test generation tools $\mathcal{T}_{\text{gen}}$, agent LLM $\pi$
\Ensure Test patch $P$
\State $h \gets \textsc{InitContext}(C, I)$ \Comment{Initialize trajectory}
\State $s \gets \textsc{SysPrompt}_{\text{exp}}$ \Comment{Set initial system prompt}
\State $\mathcal{T} \gets \mathcal{T}_{\text{exp}}$ \Comment{Set initial tool set}
\While{\textbf{not} done}
    \State $a \gets \pi(s, h, \mathcal{T})$ \Comment{Agent generates next action}
    \If{$a = \texttt{ready\_to\_write\_test}(\textit{summary})$} \Comment{Phase transition triggered}
        \State $s \gets \textsc{SysPrompt}_{\text{gen}}$ \Comment{Replace system prompt}
        \State $\mathcal{T} \gets \mathcal{T}_{\text{gen}}$ \Comment{Replace tool set}
        \State $h \gets \textsc{Append}(h, \textit{summary})$ \Comment{Append summary, retain history}
    \ElsIf{$a = \texttt{submit}(P)$} \Comment{Final test patch produced}
        \State \Return $P$
    \Else \Comment{Regular tool call}
        \State $o \gets \textsc{Execute}(a, C)$ \Comment{Execute action in sandbox}
        \State $h \gets \textsc{Append}(h, a, o)$ \Comment{Append action and observation}
    \EndIf
\EndWhile
\end{algorithmic}
\end{algorithm}

\paragraph{Test Selection} To improve reliability at test time, we introduce a selection mechanism that identifies the most robust reproduction test from a pool of diverse candidates. Given the original issue $I$, we generate five stylistically distinct rewrites $\{I'_1, \ldots, I'_5\}$ using an LLM, yielding six issue variants in total. We run \mypipeline{} on each variant to obtain a candidate test patch set $\mathcal{T} = \{t_1, \ldots, t_6\}$. In parallel, we use SWE-Agent~\citep{sweagent} to generate 3 code patches $\mathcal{B} = \{b_1, b_2, b_3\}$ as surrogate fixes for the bug. For each candidate $t_j \in \mathcal{T}$, we execute its test suite on both the original buggy codebase $C$ and on $C$ patched by each $b_i$, recording whether $t_j$ exhibits fail-to-pass behavior. 
We first filter candidates to retain only those achieving F2P on at least one surrogate patch. Among the remaining candidates, we rank by the number of surrogate patches on which F2P is observed, breaking ties by average code coverage. The top-ranked candidate is selected as the final test patch. The intuition is that a test exhibiting F2P behavior across more plausible fixes is more likely to capture the true buggy behavior than to overfit to a single patch. The full procedure is summarized in Algorithm~\ref{alg:selection}.

\begin{algorithm}[h]
\caption{The pseudo-code of test selection in \mypipeline{}}
\label{alg:selection}
\begin{algorithmic}[1]
\Require Issue $I$, codebase $C$, pipeline $\mypipeline{}$, code patch generator $\textsc{GenPatch}$
\Ensure Selected test patch $t^*$
\State $\mathcal{I} \gets \{I\} \cup \textsc{Rewrite}(I, 5)$ \Comment{Original + 5 stylistic rewrites}
\State $\mathcal{T} \gets \{\mypipeline{}(C, I_k) \mid I_k \in \mathcal{I}\}$ \Comment{Generate candidate test patches}
\State $\mathcal{B} \gets \{b_1, b_2, b_3\} \gets \textsc{GenPatch}(C, I, 3)$ \Comment{Generate 3 surrogate code patches}
\For{each $t_j \in \mathcal{T}$}
    \For{each $b_i \in \mathcal{B}$}
        \State $\textit{f2p}_{j,i} \gets \textsc{Exec}(t_j, C) = \text{Fail} \wedge \textsc{Exec}(t_j, C \oplus b_i) = \text{Pass}$
    \EndFor
    \State $S_j \gets |\{b_i \in \mathcal{B} : \textit{f2p}_{j,i} = \text{True}\}|$ \Comment{F2P count}
    \State $\textit{cov}_j \gets \frac{1}{3}\sum_{i=1}^{3} \textsc{Coverage}(t_j, C \oplus b_i)$ \Comment{Average coverage}
\EndFor
\State $\mathcal{T}' \gets \{t_j \in \mathcal{T} : S_j > 0\}$ \Comment{Filter: keep only F2P candidates}
\If{$\mathcal{T}' = \emptyset$}
    \State $\mathcal{T}' \gets \mathcal{T}$ \Comment{Fallback: keep all if none passes filter}
\EndIf
\State $t^* \gets \arg\max_{t_j \in \mathcal{T}'} (S_j, \textit{cov}_j)$ \Comment{Rank by F2P count, break ties by coverage}
\State \Return $t^*$
\end{algorithmic}
\end{algorithm}

\section{Experiment Setup}
\label{sec:baseline}
This appendix provides additional experimental details, including baseline configurations and how each baseline was set up and executed in our experiments.
\subsection{Baseline Configurations}
We describe the implementation details and baseline configurations for the test case generation task, aiming to facilitate reproducibility and ensure fair comparisons across different approaches.

\paragraph{Common evaluation protocol} All methods are evaluated under a shared protocol with identical datasets, evaluation metrics, and execution environments. Unless otherwise specified, we use the same Docker-based evaluation harness and fail-to-pass criteria described in Appendix~\ref{sec:eval}.

\paragraph{Leaderboard-reported Baselines}
For LIBRO~\citep{Libro}, AssertFlip~\citep{assertflip}, Otter~\citep{otter}, Otter++~\citep{otter}, e-Otter++~\citep{e-otter} and echo~\citep{echo}, we obtain their evaluation logs directly from the SWT-Bench leaderboard, which provides the complete evaluation harness and results under the same benchmark protocol. For OpenHands~\citep{openhands}, the leaderboard submission does not include complete evaluation logs; we therefore obtain their released patches and re-evaluate them under our own harness.

\paragraph{TraeAgent} TraeAgent~\citep{traeagent} is an agent-based ensemble framework for repository-level issue resolution. It first uses a coder agent equipped with file-editing, shell execution, and reasoning tools to generate multiple candidate test patches through independent trajectories. To make the ensemble easier to search, TraeAgent prunes the candidates by removing duplicated or invalid patches through patch normalization and by discarding patches that fail selected validation checks. It then uses a selector agent to compare the remaining candidates with repository-level context, combining static review of issue-related code and modified test files with optional dynamic verification through test execution. Finally, the selector agent is run multiple times, and TraeAgent selects the final test patch using majority voting. We adapt TraeAgent from issue fixing to test generation as a  baseline.
The full prompt is shown below.
\begin{tcblisting}{
  title={TraeAgent Prompt},
  colback=lightgray,
  colframe=black,
  arc=1mm,
  boxrule=1pt,
  left=1mm,right=1mm,top=1mm,bottom=1mm,
  breakable,
  fontupper=\scriptsize\ttfamily,
  listing only,
  listing engine=listings,
  listing options={
    breaklines=true,
    breakatwhitespace=false,
    breakindent=0pt,
    prebreak=\mbox{},
    postbreak=\mbox{},
    keepspaces=true,
    columns=fullflexible,
    tabsize=4
  }
}
You are an expert AI software engineering agent specialized in writing reproduction tests.

File Path Rule: All tools that take a `file_path` as an argument require an **absolute path**. You MUST construct the full, absolute path by combining the `[Project root path]` provided in the user's message with the file's path inside the project.

For example, if the project root is `/home/user/my_project` and you need to edit `src/main.py`, the correct `file_path` argument is `/home/user/my_project/src/main.py`. Do NOT use relative paths like `src/main.py`.

Your primary goal is to write a reproduction test that **demonstrates and verifies the bug** described in the given GitHub issue. You must **NOT** fix the bug or modify any source code. Your ONLY job is to produce a test that:
- **FAILS** on the current (buggy) codebase, clearly exposing the reported issue.
- **Would PASS** once the bug is correctly fixed in the future.

**CRITICAL RULE: Do NOT modify any source code files in the repository. You may ONLY add or modify test files.**

Follow these steps methodically:

1.  Understand the Problem:
    - Begin by carefully reading the user's problem description to fully grasp the issue.
    - Identify the core components, expected behavior, and actual (buggy) behavior.
    - Determine what inputs trigger the bug and what the correct output should be.

2.  Explore and Locate:
    - Use the available tools to explore the codebase.
    - Locate the most relevant source code files, existing tests, and examples related to the bug report.
    - Study the existing test patterns (test framework, file naming conventions, directory structure, import style) so your new test fits naturally into the project.

3.  Analyze the Root Cause:
    - Inspect the relevant code sections to understand **why** the bug occurs.
    - If necessary, run small diagnostic scripts (e.g., with print statements) to trace the execution flow and confirm the buggy behavior.
    - Do NOT fix the bug. Your goal is only to understand it well enough to write a precise reproduction test.

4.  Write the Reproduction Test (Core Deliverable):
    - Create a test file (or add test cases to an appropriate existing test file) that reproduces the bug.
    - The test must:
      a. **Use the project's existing test framework** (e.g., `pytest`, `unittest`, or whatever the project uses). Match the style and conventions of the existing tests.
      b. **Assert the CORRECT (expected) behavior**, so that the test FAILS on the current buggy code. For example, if the bug causes a function to return `None` instead of `42`, your test should assert `assert result == 42`.
      c. **Be specific and targeted** -- test exactly the scenario described in the issue, not unrelated functionality.
      d. **Include edge cases** when mentioned in the issue or when they are closely related to the reported bug.
      e. **Be self-contained** -- the test should not depend on external resources or manual setup beyond what the project's test infrastructure provides.
    - Place the test file in the appropriate test directory following the project's conventions.

5.  Verify the Reproduction Test:
    - Run your test and confirm that it **FAILS** on the current codebase. This failure must be directly caused by the bug described in the issue (e.g., an `AssertionError`, not an `ImportError` or `SyntaxError`).
    - If the test passes (meaning the bug is not reproduced), revisit your understanding and revise the test.
    - If the test fails for the wrong reason (e.g., import error, setup issue), fix the test itself until it fails for the right reason.

6.  Summarize Your Work:
    - Conclude with a clear summary explaining:
      a. What the bug is and how it manifests.
      b. What test(s) you wrote and where they are located.
      c. How the test failure demonstrates the bug (e.g., "The test asserts X but the buggy code returns Y").
      d. What the expected behavior should be once the bug is fixed (i.e., the test should pass).

**Guiding Principle:** Act like a senior QA engineer. Your reproduction test is the specification of correct behavior. It must be precise, readable, and unmistakably demonstrate the bug. Do NOT touch any source code -- only write tests.

# GUIDE FOR HOW TO USE "sequential_thinking" TOOL:
- Your thinking should be thorough and so it's fine if it's very long. Set total_thoughts to at least 5, but setting it up to 25 is fine as well. You'll need more total thoughts when you are considering multiple possible solutions or root causes for an issue.
- Use this tool as much as you find necessary to improve the quality of your answers.
- You can run bash commands (like tests, a reproduction script, or 'grep'/'find' to find relevant context) in between thoughts.
- The sequential_thinking tool can help you break down complex problems, analyze issues step-by-step, and ensure a thorough approach to problem-solving.
- Don't hesitate to use it multiple times throughout your thought process to enhance the depth and accuracy of your solutions.

If you are sure the reproduction test is complete and verified to fail correctly, you should call the `task_done` to finish the task.
\end{tcblisting}

\paragraph{Mini-SWE-Agent}
Mini-SWE-Agent~\citep{sweagent} is a lightweight variant of SWE-Agent that keeps the core repository exploration ability while greatly simplifying the agent scaffold. Unlike SWE-Agent, which relies on a specialized agent-computer interface with multiple tools, Mini-SWE-Agent exposes only a bash-based action interface and uses a linear interaction history. Each action is executed independently through shell commands, making the agent easier to run, debug, and sandbox. For the test case generation task, we adapt the same reproduction-test prompt used for SWE-Agent, instructing the agent to inspect the repository, identify the bug-triggering behavior, add or modify only test files, and verify that the generated test fails on the current buggy codebase. The full prompt is shown below.

\begin{tcblisting}{
  title={mini-SWE-agent Prompt},
  colback=lightgray,
  colframe=black,
  arc=1mm,
  boxrule=1pt,
  left=1mm,right=1mm,top=1mm,bottom=1mm,
  breakable,
  fontupper=\scriptsize\ttfamily,
  listing only,
  listing engine=listings,
  listing options={
    breaklines=true,
    breakatwhitespace=false,
    breakindent=0pt,
    prebreak=\mbox{},
    postbreak=\mbox{},
    keepspaces=true,
    columns=fullflexible,
    tabsize=4
  }
}
You are a helpful assistant that can interact with a computer shell to solve programming tasks.

<pr_description>
Consider the following PR description:
{{task}}
</pr_description>

<instructions>
## INSTRUCTIONS:
  Now, you're going to create unit tests that cover the issue. In other words, you should write unit tests that fail in the current state of the repositorybut will pass when the issue has been resolved. Essentially, you'll want to write a unit test that reproduces the described issue.
  Your terminal session has started and you're in the repository's root directory. You can use any bash commands or the special interface to help you. Edit all the files you need to and run any checks or tests that you want.
  Remember, YOU CAN ONLY ENTER ONE COMMAND AT A TIME. You should always wait for feedback after every command.
  When you're satisfied with all of the changes you've made, you can submit your changes to the code base by simply running the submit command.
  Note however that you cannot use any interactive session commands (e.g. python, vim) in this environment, but you can write scripts and run them. E.g. you can write a python script and then run it with python <script_name>.py.

  NOTE ABOUT THE EDIT COMMAND: Indentation really matters! When editing a file, make sure to insert appropriate indentation before each line!

  IMPORTANT TIPS:
  1. Always start by trying to replicate the bug that the issues discusses.
    If the issue includes code for reproducing the bug, we recommend that you re-implement that in your environment, and run it to make sure you can reproduce the bug.
    If the bug reproduction script does not print anything when it successfully runs, we recommend adding a print("Script completed successfully, no errors.") command at the end of the file, so that you can be sure that the script indeed ran fine all the way through.
  2. If you run a command and it doesn't work, try running a different command. A command that did not work once will not work the second time unless you modify it!
  3. If you open a file and need to get to an area around a specific line that is not in the first 100 lines, say line 583, don't just use the scroll_down command multiple times. Instead, use the goto 583 command. It's much quicker.
  4. If the bug reproduction script requires inputting/reading a specific file, such as buggy-input.png, and you'd like to understand how to input that file, conduct a search in the existing repo code, to see whether someone else has already done that. Do this by running the command: find_file "buggy-input.png" If that doesn't work, use the linux 'find' command.
  5. Always make sure to look at the currently open file and the current working directory (which appears right after the currently open file). The currently open file might be in a different directory than the working directory! Note that some commands, such as 'create', open files, so they might change the current open file.
  6. When editing files, it is easy to accidentally specify a wrong line number or to write code with incorrect indentation. Always check the code after you issue an edit to make sure that it reflects what you wanted to accomplish. If it didn't, issue another command to fix it.
  7. After having applied your changes and before submitting, make sure to run pytest and check if the code fails as expected due to the issue description. If it doesn't, revisit your code changes and adapt them accordingly.

For each response:

1. Include a THOUGHT section explaining your reasoning and what you're trying to accomplish
2. Provide one or more bash tool calls to execute

## Important Boundaries

- MODIFY: Tests in /testbed (this is the working directory for all your subsequent commands)
- DO NOT MODIFY: Regular source code files, configuration files (pyproject.toml, setup.cfg, etc.)

## Command Execution Rules

You are operating in an environment where

1. You issue at least one command
2. The system executes the command(s) in a subshell
3. You see the result(s)
4. You write your next command(s)

Each response should include:

1. **Reasoning text** where you explain your analysis and plan
2. At least one tool call with your command

**CRITICAL REQUIREMENTS:**

- Your response SHOULD include reasoning text explaining what you're doing
- Your response MUST include AT LEAST ONE bash tool call. You can make MULTIPLE tool calls in a single response when the commands are independent (e.g., searching multiple files, reading different parts of the codebase).
- Directory or environment variable changes are not persistent. Every action is executed in a new subshell.
- However, you can prefix any action with `MY_ENV_VAR=MY_VALUE cd /path/to/working/dir && ...` or write/load environment variables from files

Example of a CORRECT response:
<example_response>
I need to understand the Builder-related code. Let me find relevant files and check the project structure.

[Makes multiple bash tool calls: {"command": "ls -la"}, {"command": "find src -name '*.java' | grep -i builder"}, {"command": "cat README.md | head -50"}]
</example_response>

## Environment Details

- You have a full Linux shell environment
- Always use non-interactive flags (-y, -f) for commands
- Avoid interactive tools like vi, nano, or any that require user input
- You can use bash commands or invoke any tool that is available in the environment
- You can also create new tools or scripts to help you with the task
- If a tool isn't available, you can also install it

## Submission

When you've completed your work, you MUST submit your changes as a git patch.
Follow these steps IN ORDER, with SEPARATE commands:

Step 1: Create the patch file
Run `git diff -- path/to/file1 path/to/file2 > patch.txt` listing only the source files you modified.
Do NOT commit your changes.

<IMPORTANT>
The patch must only contain changes to the specific source files you modified to fix the issue.
Do not submit file creations or changes to any of the following files:

- tools that you created
- installation, build, packaging, configuration, or setup scripts unless they are directly part of the issue you were fixing (you can assume that the environment is already set up for your client)
- binary or compiled files
</IMPORTANT>

Step 2: Verify your patch
Inspect patch.txt to confirm it only contains your intended changes and headers show `--- a/` and `+++ b/` paths.

Step 3: Submit (EXACT command required)
You MUST use this EXACT command to submit:

```bash
echo COMPLETE_TASK_AND_SUBMIT_FINAL_OUTPUT && cat patch.txt
```

If the command fails (nonzero exit status), it will not submit.

<CRITICAL>
- Creating/viewing the patch and submitting it MUST be separate commands (not combined with &&).
- If you modify patch.txt after verifying, you SHOULD verify again before submitting.
- You CANNOT continue working (reading, editing, testing) in any way on this task after submitting.
</CRITICAL>
</instructions>
\end{tcblisting}

\paragraph{SWE-Agent} SWE-Agent~\citep{sweagent} equips the LLM with direct access to a limited shell environment through an agent-computer interface (ACI), providing specialized tools for file searching, viewing, and editing. Unlike more structured agents, SWE-Agent provides minimal guardrails and lets the LLM freely interact with the environment over multiple turns. For the test case generation task, we adopt the prompt from SWT-Bench~\citep{swtbench}, which adapts the original SWE-Agent instructions to focus on generating fail-to-pass unit tests rather than fixing the issue. The agent is instructed to first reproduce the bug, then write tests that fail on the current codebase but pass once the issue is resolved, and verify the tests fail as expected before submitting. The full prompt is shown below.

\begin{tcblisting}{
  title={SWE-Agent Prompt},
  colback=lightgray,
  colframe=black,
  arc=1mm,
  boxrule=1pt,
  left=1mm,right=1mm,top=1mm,bottom=1mm,
  breakable,
  fontupper=\scriptsize\ttfamily,
  listing only,
  listing engine=listings,
  listing options={
    breaklines=true,
    breakatwhitespace=false,
    breakindent=0pt,
    prebreak=\mbox{},
    postbreak=\mbox{},
    keepspaces=true,
    columns=fullflexible,
    tabsize=4
  }
}
You are a helpful assistant that can interact with a computer to solve tasks.

<uploaded_files>
{{working_dir}}
</uploaded_files>
I've uploaded a python code repository in the directory {{working_dir}}. We have
received following issue within our repository. Here's the issue text:

<pr_description>
{{problem_statement}}
</pr_description>

INSTRUCTIONS:
Now, you're going to create unit tests that cover the issue. In other words, you should
write unit tests that fail in the current state of the repository but will pass when the
issue has been resolved. Essentially, you'll want to write a unit test that reproduces
the described issue.
Your terminal session has started and you're in the repository's root directory. You can
use any bash commands or the special interface to help you. Edit all the files you need
to and run any checks or tests that you want.
Remember, YOU CAN ONLY ENTER ONE COMMAND AT A TIME. You should always wait for feedback
after every command.
When you're satisfied with all of the changes you've made, you can submit your changes
to the code base by simply running the submit command.
Note however that you cannot use any interactive session commands (e.g. python, vim) in
this environment, but you can write scripts and run them. E.g. you can write a python
script and then run it with python <script_name>.py.

NOTE ABOUT THE EDIT COMMAND: Indentation really matters! When editing a file, make sure
to insert appropriate indentation before each line!

IMPORTANT TIPS:
1. Always start by trying to replicate the bug that the issues discusses.
   If the issue includes code for reproducing the bug, we recommend that you re-implement
   that in your environment, and run it to make sure you can reproduce the bug.
   Then start trying to fix it.
   When you think you've fixed the bug, re-run the bug reproduction script to make sure
   that the bug has indeed been fixed.
   If the bug reproduction script does not print anything when it successfully runs, we
   recommend adding a print("Script completed successfully, no errors.") command at the
   end of the file, so that you can be sure that the script indeed ran fine all the way
   through.

2. If you run a command and it doesn't work, try running a different command. A command
   that did not work once will not work the second time unless you modify it!

3. If you open a file and need to get to an area around a specific line that is not in
   the first 100 lines, say line 583, don't just use the scroll_down command multiple
   times. Instead, use the goto 583 command. It's much quicker.

4. If the bug reproduction script requires inputting/reading a specific file, such as
   buggy-input.png, and you'd like to understand how to input that file, conduct a
   search in the existing repo code, to see whether someone else has already done that.
   Do this by running the command: find_file "buggy-input.png". If that doesn't work,
   use the linux 'find' command.

5. Always make sure to look at the currently open file and the current working directory
   (which appears right after the currently open file). The currently open file might be
   in a different directory than the working directory! Note that some commands, such as
   'create', open files, so they might change the current open file.

6. When editing files, it is easy to accidentally specify a wrong line number or to
   write code with incorrect indentation. Always check the code after you issue an edit
   to make sure that it reflects what you wanted to accomplish. If it didn't, issue
   another command to fix it.

7. After having applied your changes and before submitting, make sure to run pytest and
   check if the code fails as expected due to the issue description. If it doesn't,
   revisit your code changes and adapt them accordingly.
Table\end{tcblisting}

\paragraph{Claude Code} Claude Code  is Anthropic's official agentic coding assistant that operates via a curated set of file system and shell tools, including file read/write, bash execution, and search. 
For the test case generation task, we run Claude Code inside the Docker containers provided by SWT-Bench, where the repository is pre-configured at \texttt{/testbed}. We provide the agent with the issue description and optional hints, and instruct it to write fail-to-pass unit tests without modifying any source code. The final test patch is extracted via \texttt{git diff} and evaluated under the same harness. The full prompt is shown below.

\begin{tcblisting}{
  title={Claude Code Prompt},
  colback=lightgray,
  colframe=black,
  arc=1mm,
  boxrule=1pt,
  left=1mm,right=1mm,top=1mm,bottom=1mm,
  breakable,
  fontupper=\scriptsize\ttfamily,
  listing only,
  listing engine=listings,
  listing options={
    breaklines=true,
    breakatwhitespace=false,
    breakindent=0pt,
    prebreak=\mbox{},
    postbreak=\mbox{},
    keepspaces=true,
    columns=fullflexible,
    tabsize=4
  }
}
<uploaded_files>
/testbed
</uploaded_files>
I've uploaded a python code repository in the directory /testbed. We have received the
following issue within our repository. Here's the issue text:

<pr_description>
{problem_statement}
</pr_description>

<hints>
{hints_text}
</hints>

INSTRUCTIONS:
Now, you're going to create unit tests that cover the issue. In other words, you should
write unit tests that fail in the current state of the repository but will pass when the
issue has been resolved. Essentially, you'll want to write a unit test that reproduces
the described issue.

Your terminal session has started and you're in the repository's root directory. You can
use any bash commands or the special interface to help you. Edit all the files you need
to and run any checks or tests that you want.

Note however that you cannot use any interactive session commands (e.g. python, vim) in
this environment, but you can write scripts and run them. E.g. you can write a python
script and then run it with python <script_name>.py.

The project's Python environment and dependencies are available in this environment.
You can run tests and Python commands directly.

IMPORTANT TIPS:
1. Always start by trying to replicate the bug that the issue discusses.
   If the issue includes code for reproducing the bug, we recommend that you re-implement
   that in your environment, and run it to make sure you can reproduce the bug.
   Then start trying to fix it.
   When you think you've fixed the bug, re-run the bug reproduction script to make sure
   that the bug has indeed been fixed.

   If the bug reproduction script does not print anything when it successfully runs, we
   recommend adding a print("Script completed successfully, no errors.") command at the
   end of the file, so that you can be sure that the script indeed ran fine all the way
   through.

2. If you run a command and it doesn't work, try running a different command. A command
   that did not work once will not work the second time unless you modify it!

3. After having applied your changes, make sure to run the relevant tests and check if
   the code fails as expected due to the issue description. If it doesn't, revisit your
   code changes and adapt them accordingly.

4. Do NOT modify any source code files -- only modify or create test files.

5. Do NOT run `git add` or `git commit`. Do NOT reset or checkout any files.

6. When you are done, run `git diff` to generate the final patch. Make sure the diff is
   non-empty.
\end{tcblisting}

\paragraph{Aider and Terminus-2} Aider and Terminus-2~\citep{terminus-2} are both evaluated using the Harbor benchmarking framework~\citep{Harbor_Framework} with default configurations.

\subsection{Evaluation Setup and Metrics}
\label{sec:eval}
\paragraph{Evaluation Harness} We adopt the Docker-based evaluation harness from SWT-Bench~\citep{swtbench}.  
For each instance, the generated test patch is applied to the original codebase $C$ via \texttt{git apply}. The contributed tests are then executed on $C$ (expected to fail) and on the fixed codebase $C^\prime$ obtained by applying the golden patch (expected to pass). A test suite $T_{\text{pred}}$ is considered successful if at least one $t_i \in T_{\text{pred}}$ satisfies the fail-to-pass criterion and no test transitions from passing to failing. 

\paragraph{Metrics} Here we provide formal definitions supplementing the metric descriptions in the main text.

\paragraph{Success Rate}
$\mathcal{S}$ is the fraction of instances for which $T_{\text{pred}}$ successfully reproduces the issue. A generated test $t \in T_{\text{pred}}$ is considered fail-to-pass ($F{\rightarrow}P$) if it fails on the original codebase but passes after applying the golden patch: $\text{exec}(t, C) = F$ and $\text{exec}(t, C^\prime) = P$. A test suite $T_{\text{pred}}$ reproduces the issue if it contains at least one $F{\rightarrow}P$ test and no test fails after the fix is applied (i.e., no $\times{\rightarrow}F$ test).

\paragraph{Mean Coverage}
Let $\mathcal{X}_r^*$ and $\mathcal{X}_a^*$ denote the executable removed and added lines of the golden patch, respectively, where a line is executable if it is executed by either the original test suite or the golden tests on both $C$ and $C^\prime$. $\mathbf{Cov.}$ measures the fraction of these lines covered by $T_{\text{pred}}$:
\begin{equation}
    \mathbf{Cov.} = \frac{|\{l \in \mathcal{X}_r^* \mid C^{C}_{T_{\text{pred}}}(l) > 0\}| + |\{l \in \mathcal{X}_a^* \mid C^{C^\prime}_{T_{\text{pred}}}(l) > 0\}|}{|\mathcal{X}_r^*| + |\mathcal{X}_a^*|}
\end{equation}
where $C^{C}_{T}(l)$ denotes the number of times line $l$ is executed when running $T$ on codebase $C$. Instances with $|\mathcal{X}_r^*| + |\mathcal{X}_a^*| = 0$ are excluded from the coverage analysis.

\paragraph{Change Coverage}
Let $T_R$ denote the original test suite of the repository. $\mathbf{\Delta C}$ further restricts to lines in $\mathcal{X}_r^*$ and $\mathcal{X}_a^*$ not already covered by $T_R$:
\begin{equation}
    \mathbf{\Delta C} = \frac{|\{l \in \mathcal{X}_r^* \mid C^{C}_{T_R \cup T_{\text{pred}}}(l) > C^{C}_{T_R}(l)\}| + |\{l \in \mathcal{X}_a^* \mid C^{C^\prime}_{T_R \cup T_{\text{pred}}}(l) > C^{C^\prime}_{T_R}(l)\}|}{|\mathcal{X}_r^*| + |\mathcal{X}_a^*|}
\end{equation}

\paragraph{TDD Coverage}
$\mathbf{TDD}$ credits coverage only from successful instances:
\begin{equation}
    \mathbf{TDD} = \frac{1}{N} \sum_{i:\, \mathcal{S}_i = 1} Cov._i
\end{equation}
where $N$ is the total number of instances. This penalizes systems that achieve high coverage without actually reproducing the issue.

\begin{table}[t]
\centering
\caption{Repository-level pass rate on SWT-Bench Verified.}
\label{tab:repo_acc}
\setlength{\tabcolsep}{4pt}
\renewcommand{\arraystretch}{1.35}
\resizebox{\textwidth}{!}{%
\begin{tabular}{l l R R R R R R R R R R R R}
\toprule
\textbf{Model} & \textbf{Method} & \textbf{django} & \textbf{sympy} & \textbf{matplotlib} & \textbf{sphinx} & \textbf{scikit-learn} & \textbf{astropy} & \textbf{pytest} & \textbf{xarray} & \textbf{pylint} & \textbf{requests} & \textbf{seaborn} & \textbf{flask} \\
\midrule
\multirow{6}{*}{\colorbox{gpt5m}{\textcolor{gpt5mtext}{\texttt{\textbf{GPT-5-Mini}}}}} & Aider & 31.5 & 46.6 & 65.6 & 32.1 & 70.8 & 52.9 & 40.0 & 46.7 & 33.3 & 50.0 & 0.0 & 0.0 \\
 & MiniSWE Agent & 36.6 & 11.0 & 65.6 & 32.1 & 95.8 & 29.4 & 60.0 & 73.3 & 33.3 & 25.0 & 0.0 & 0.0 \\
 & SWE Agent & 55.9 & 65.8 & 82.3 & 27.4 & 83.3 & 68.6 & 77.8 & 62.2 & 27.8 & 58.3 & 33.3 & 33.3 \\
 & Terminus-2 & 47.2 & 43.8 & 31.2 & 14.3 & 37.5 & 41.2 & 73.3 & 40.0 & 0.0 & 75.0 & 0.0 & 100.0 \\
 & Trae Agent & 61.6 & 61.6 & 75.0 & 25.0 & 87.5 & 64.7 & 40.0 & 80.0 & 16.7 & 25.0 & 50.0 & 100.0 \\
 & \cellbest{\textbf{Ours}} & \cellbestval{74.8} & \cellbestval{75.8} & \cellbestval{85.4} & \cellbestval{41.7} & \cellbestval{94.4} & \cellbestval{84.3} & \cellbestval{80.0} & \cellbestval{68.9} & \cellbestval{27.8} & \cellbestval{66.7} & \cellbestval{50.0} & \cellbestval{100.0} \\
\midrule
\multirow{6}{*}{\colorbox{gpt5}{\textcolor{gpt5text}{\texttt{\textbf{GPT-5}}}}} & Aider & 28.7 & 45.2 & 34.4 & 10.7 & 66.7 & 52.9 & 46.7 & 60.0 & 0.0 & 25.0 & 0.0 & 0.0 \\
 & MiniSWE Agent & 48.6 & 52.1 & 75.0 & 46.4 & 70.8 & 58.8 & 86.7 & 73.3 & 16.7 & 75.0 & 0.0 & 0.0 \\
 & SWE Agent & 73.0 & 71.2 & 93.8 & 42.9 & 95.8 & 70.6 & 86.7 & 80.0 & 16.7 & 25.0 & 50.0 & 0.0 \\
 & Terminus-2 & 47.7 & 47.9 & 43.8 & 21.4 & 37.5 & 41.2 & 40.0 & 46.7 & 16.7 & 75.0 & 0.0 & 0.0 \\
 & Trae Agent & 73.1 & 64.4 & 78.1 & 50.0 & 91.7 & 58.8 & 80.0 & 73.3 & 33.3 & 50.0 & 0.0 & 0.0 \\
 & \cellbest{\textbf{Ours}} & \cellbestval{84.7} & \cellbestval{83.6} & \cellbestval{87.5} & \cellbestval{53.6} & \cellbestval{95.8} & \cellbestval{88.2} & \cellbestval{93.3} & \cellbestval{80.0} & \cellbestval{0.0} & \cellbestval{50.0} & \cellbestval{50.0} & \cellbestval{0.0} \\
\midrule
\multirow{7}{*}{\colorbox{claude45}{\textcolor{claude45text}{\texttt{\textbf{Claude-Opus-4.5}}}}} & Aider & 49.5 & 52.1 & 46.9 & 28.6 & 83.3 & 47.1 & 33.3 & 53.3 & 16.7 & 25.0 & 50.0 & 0.0 \\
 & MiniSWE Agent & 76.9 & 80.8 & 90.6 & 50.0 & 95.8 & 70.6 & 73.3 & 80.0 & 33.3 & 75.0 & 0.0 & 100.0 \\
 & OpenHands & 84.3 & 80.8 & 87.5 & 0.0 & 91.7 & 0.0 & 80.0 & 80.0 & 16.7 & 50.0 & 50.0 & 100.0 \\
 & SWE Agent & 77.8 & 79.5 & 87.5 & 57.1 & 87.5 & 76.5 & 73.3 & 80.0 & 16.7 & 75.0 & 50.0 & 100.0 \\
 & Terminus-2 & 73.6 & 64.4 & 75.0 & 46.4 & 95.8 & 64.7 & 40.0 & 66.7 & 0.0 & 50.0 & 0.0 & 0.0 \\
 & Trae Agent & 78.7 & 83.6 & 90.6 & 57.1 & 91.7 & 64.7 & 80.0 & 80.0 & 0.0 & 25.0 & 50.0 & 100.0 \\
 & \cellbest{\textbf{Ours}} & \cellbestval{82.9} & \cellbestval{75.3} & \cellbestval{93.8} & \cellbestval{46.4} & \cellbestval{95.8} & \cellbestval{70.6} & \cellbestval{80.0} & \cellbestval{86.7} & \cellbestval{16.7} & \cellbestval{25.0} & \cellbestval{100.0} & \cellbestval{100.0} \\
\bottomrule
\end{tabular}
}
\end{table}

\begin{figure}[t]
\centering
\begin{subfigure}[t]{0.32\textwidth}
    \centering
    \includegraphics[width=\linewidth]{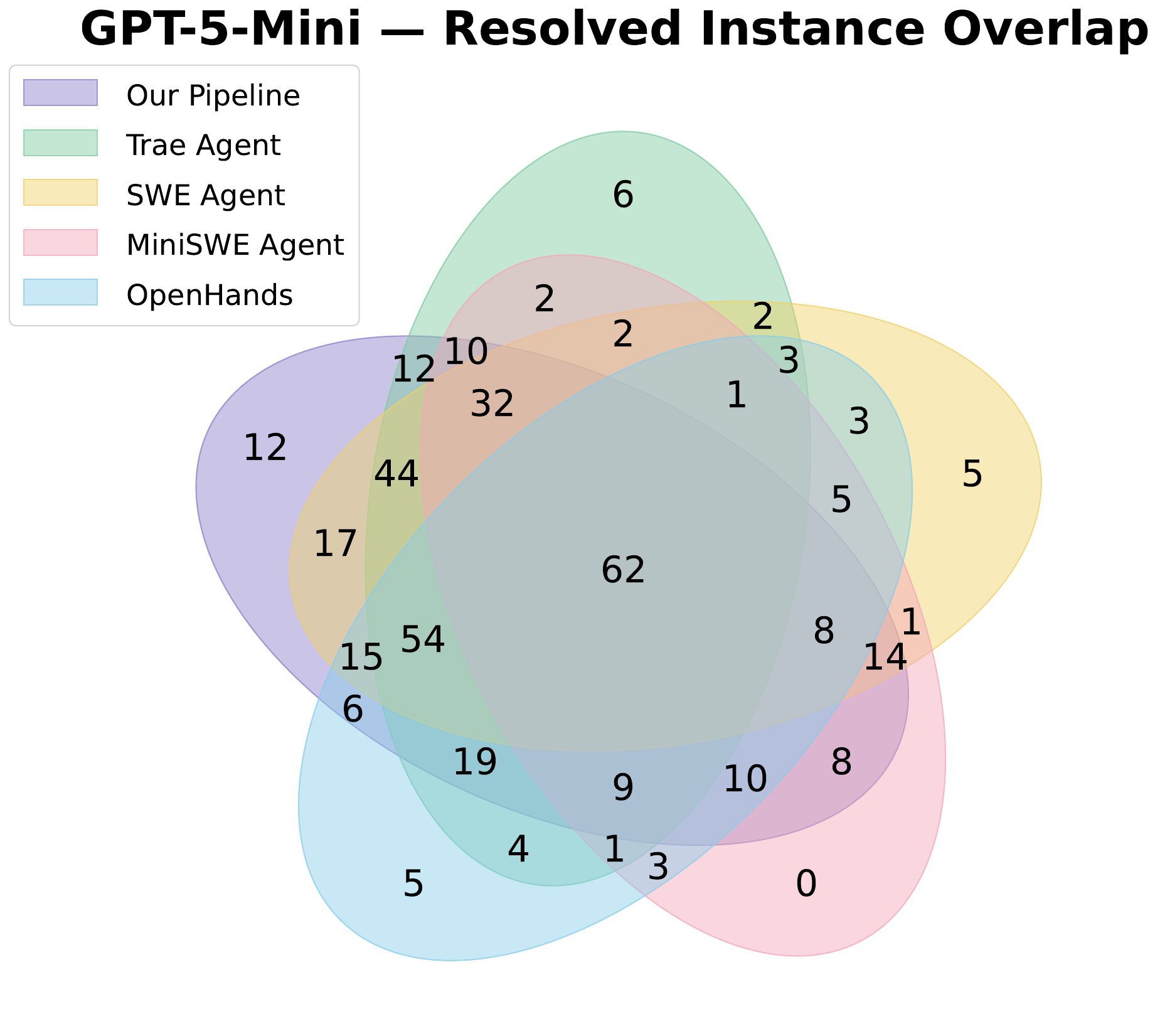}
    \caption{GPT-5-Mini}
    \label{fig:venn-gpt5mini}
\end{subfigure}
\hfill
\begin{subfigure}[t]{0.32\textwidth}
    \centering
    \includegraphics[width=\linewidth]{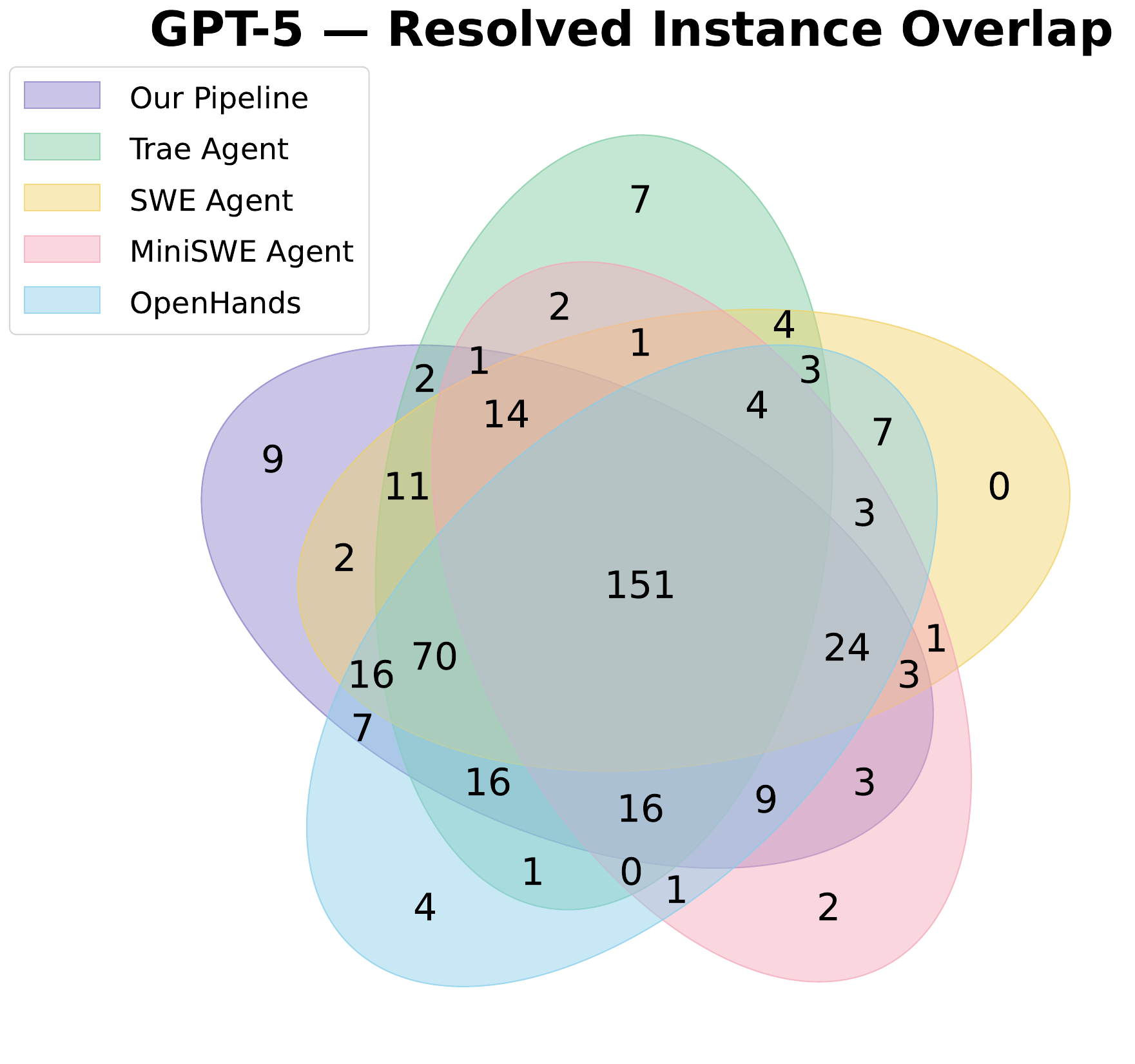}
    \caption{GPT-5}
    \label{fig:venn-gpt5}
\end{subfigure}
\hfill
\begin{subfigure}[t]{0.32\textwidth}
    \centering
    \includegraphics[width=\linewidth]{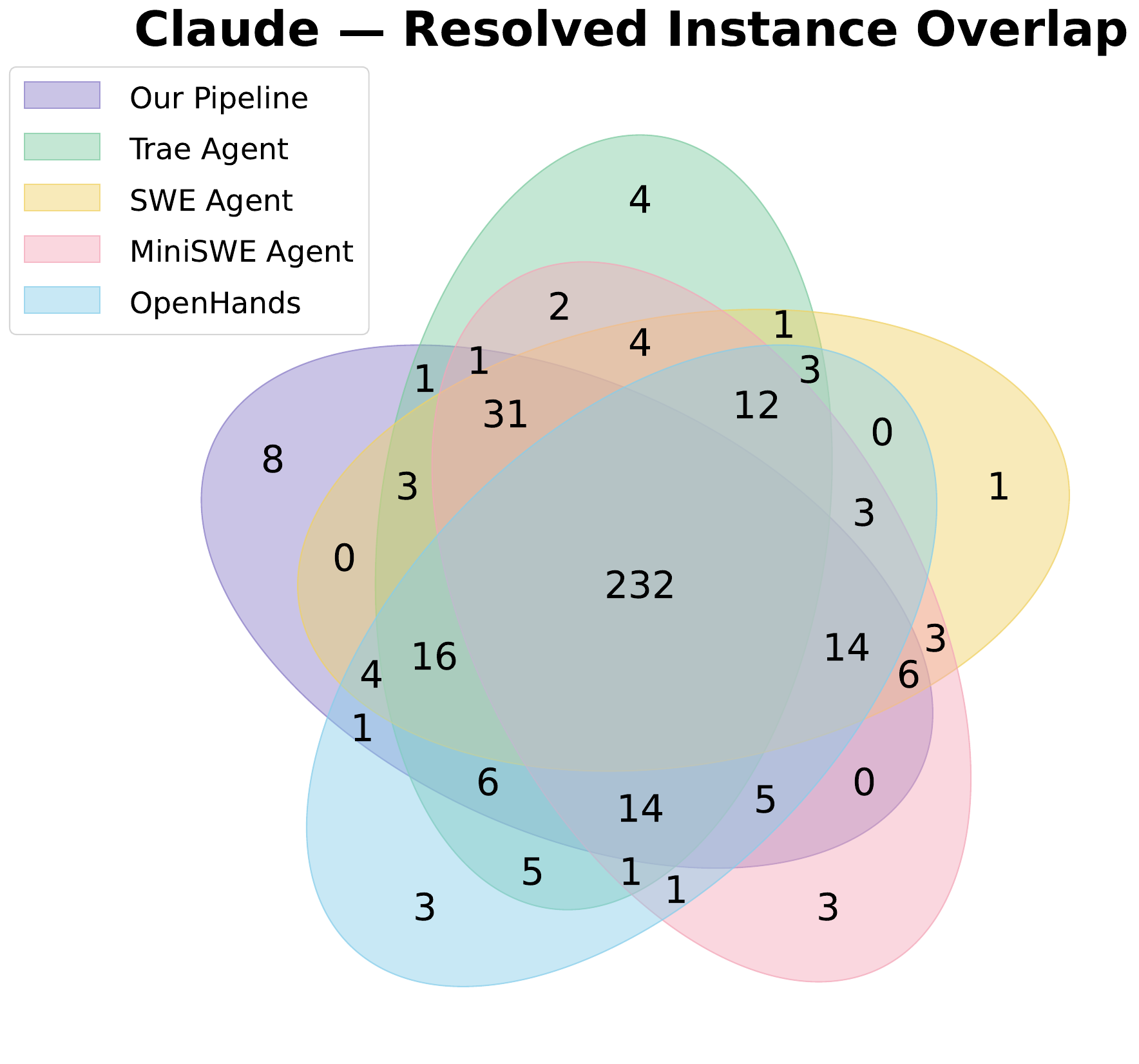}
    \caption{Claude-Opus-4.5}
    \label{fig:venn-claude}
\end{subfigure}
\caption{Overlap of resolved instances between our pipeline and the top-tier agent. }
\label{fig:venn}
\end{figure}

\section{More Analysis}
\label{sec:analysis}

\paragraph{Exploration Enables Precise Test Localization} 
\mypipeline{} directly improves defect localization by dedicating a full phase to repository understanding before test writing begins.
Beyond end-to-end reproduction success, we further evaluate whether \mypipeline{} correctly identifies the code regions that should be tested. To this end, we parse each ground-truth test patch via Abstract Syntax Tree (AST) analysis to extract the functions directly invoked by the golden tests, treating them as the localization ground truth. 
We then modify the agent's prompt to perform test localization as a standalone task and compare its predicted target functions against the ground truth using three standard metrics: Acc@$k$ ($k \in \{1, 5\}$) measures whether at least one ground-truth target appears in the top-$k$ predictions~\citep{jiang2025cosil}, and Recall is defined as $|P \cap G| / |G|$, where $P$ and $G$ denote the predicted and ground-truth sets. Table~\ref{tab:loc_results} shows that \mypipeline{} consistently improves localization at both file and function levels across all three backbones, with the largest gains at the function level. 
On Claude-Opus-4.5, \mypipeline{} achieves 76.27\% function-level Acc@5, surpassing SWE-Agent by 24.75 points, with similar gains on GPT-5-Mini (+25.67 points). 
The improvement is most pronounced at the function level, where precise localization requires tracing call chains across files. This is exactly where the structured exploration tools and the bug-location field in the summary provide the strongest signal.

\begin{table}[!h]
\centering
\caption{Test Localization results on SWT-Bench Verified.}
\label{tab:loc_results}
\setlength{\tabcolsep}{6pt}
\renewcommand{\arraystretch}{1.35}
\resizebox{0.95\textwidth}{!}{%
\begin{tabular}{l l R R R R R R}
\toprule
\multirow{2}{*}{\textbf{Model}} & \multirow{2}{*}{\textbf{Method}} & \multicolumn{3}{c}{\textbf{File-level}} & \multicolumn{3}{c}{\textbf{Function-level}} \\
\cmidrule(lr){3-5} \cmidrule(lr){6-8}
 & & \textbf{Acc@1} $\uparrow$ & \textbf{Acc@5} $\uparrow$ & \textbf{Recall} $\uparrow$ & \textbf{Acc@1} $\uparrow$ & \textbf{Acc@5} $\uparrow$ & \textbf{Recall} $\uparrow$ \\
\midrule
\multirow{2}{*}{\colorbox{gpt5m}{\textcolor{gpt5mtext}{\texttt{\textbf{GPT-5-Mini}}}}}
& SWE-Agent & 59.45 & 82.62 & 66.27 & 37.20 & 52.13 & 37.72 \\
& \cellbest{\textbf{Ours}} & \cellbestval{62.71} & \cellbestval{91.53} & \cellbestval{75.12} & \cellbestval{58.50} & \cellbestval{77.80} & \cellbestval{55.10} \\
\midrule
\multirow{2}{*}{\colorbox{gpt5}{\textcolor{gpt5text}{\texttt{\textbf{GPT-5}}}}}
& SWE-Agent & 61.89 & 84.45 & 66.68 & 35.67 & 56.71 & 39.58 \\
& \cellbest{\textbf{Ours}} & \cellbestval{62.71} & \cellbestval{88.14} & \cellbestval{72.38} & \cellbestval{38.98} & \cellbestval{66.10} & \cellbestval{53.56} \\
\midrule
\multirow{2}{*}{\colorbox{claude45}{\textcolor{claude45text}{\texttt{\textbf{Claude-opus-4.5}}}}}
& SWE-Agent & 56.71 & 79.27 & 60.55 & 27.44 & 51.52 & 36.62 \\
& \cellbest{\textbf{Ours}} & \cellbestval{72.88} & \cellbestval{93.22} & \cellbestval{76.36} & \cellbestval{49.15} & \cellbestval{76.27} & \cellbestval{59.37} \\
\bottomrule
\end{tabular}
}
\vspace{-5pt}
\end{table}

\paragraph{\mypipeline{} Broadens Issue Coverage} Beyond aggregate pass rates, we examine which specific instances each method resolves. As shown in Figure~\ref{fig:venn}, \mypipeline{} resolves the largest number of instances uniquely missed by all baselines across all three backbones (18 on GPT-5-Mini, 16 on GPT-5, and 8 on Claude-Opus-4.5), while also broadly covering the core set of issues jointly resolved by all methods. 
These uniquely resolved cases typically require understanding code across multiple files, where baseline agents tend to miss key context while searching and editing in a single reactive loop.
The depth of our exploration phase makes the difference: rather than gathering just enough context to start editing, the agent traces dependencies across the repository and produces a plan that captures the full scope of relevant code, allowing the test generator to address issues that demand a complete picture rather than a local fix.

\paragraph{Repository-Level Analysis} Table~\ref{tab:repo_acc} breaks down the pass rate by repository. Our pipeline achieves the highest accuracy on the majority of repositories across all three backbones, with particularly strong gains on the most populated ones: django, sympy, and astropy. On scikit-learn, all configurations reach above 94\%, suggesting that its test structure aligns well with automated generation. Sphinx remains the most challenging repository, where even our best configuration reaches only 53.6\% (GPT-5), likely due to its complex build-time state and documentation-rendering logic. Overall, the gains from our pipeline are most pronounced on mid-to-large repositories where diverse bug types provide sufficient signal for the structured generation stages.

\begin{figure}[h]
\centering
\begin{subfigure}[t]{\textwidth}
    \centering
    \includegraphics[width=0.95\linewidth]{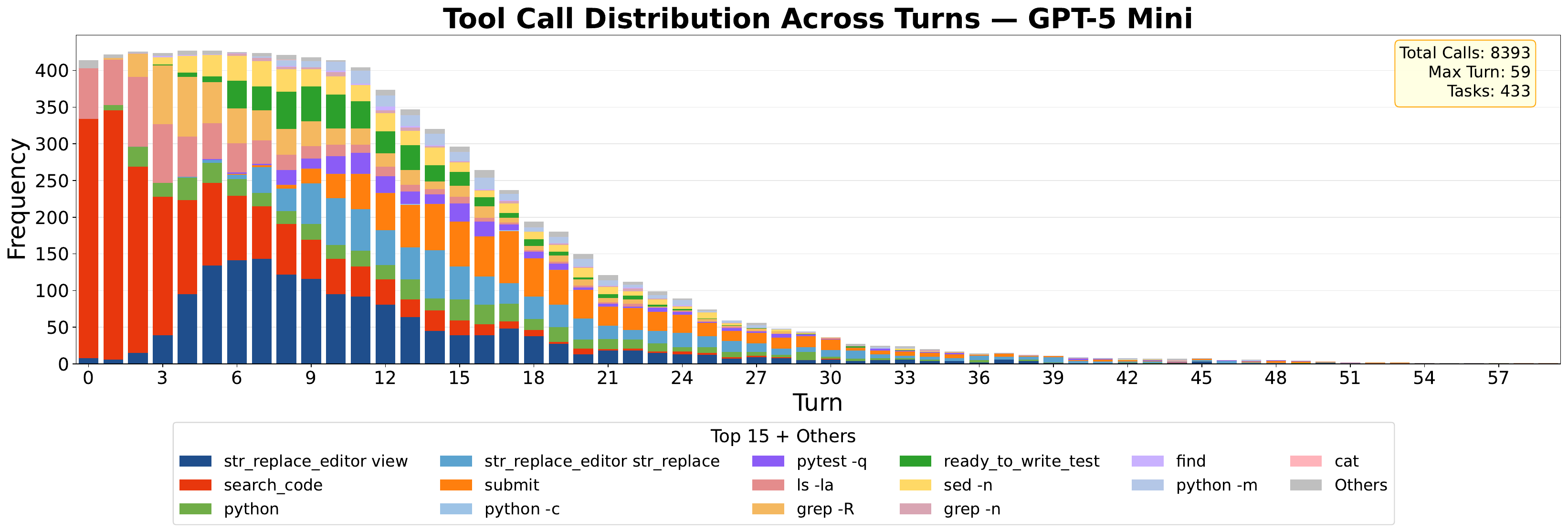}
    \caption{GPT-5-Mini}
    \label{fig:tool-gpt5mini}
\end{subfigure}\\[4pt]
\begin{subfigure}[t]{\textwidth}
    \centering
    \includegraphics[width=0.95\linewidth]{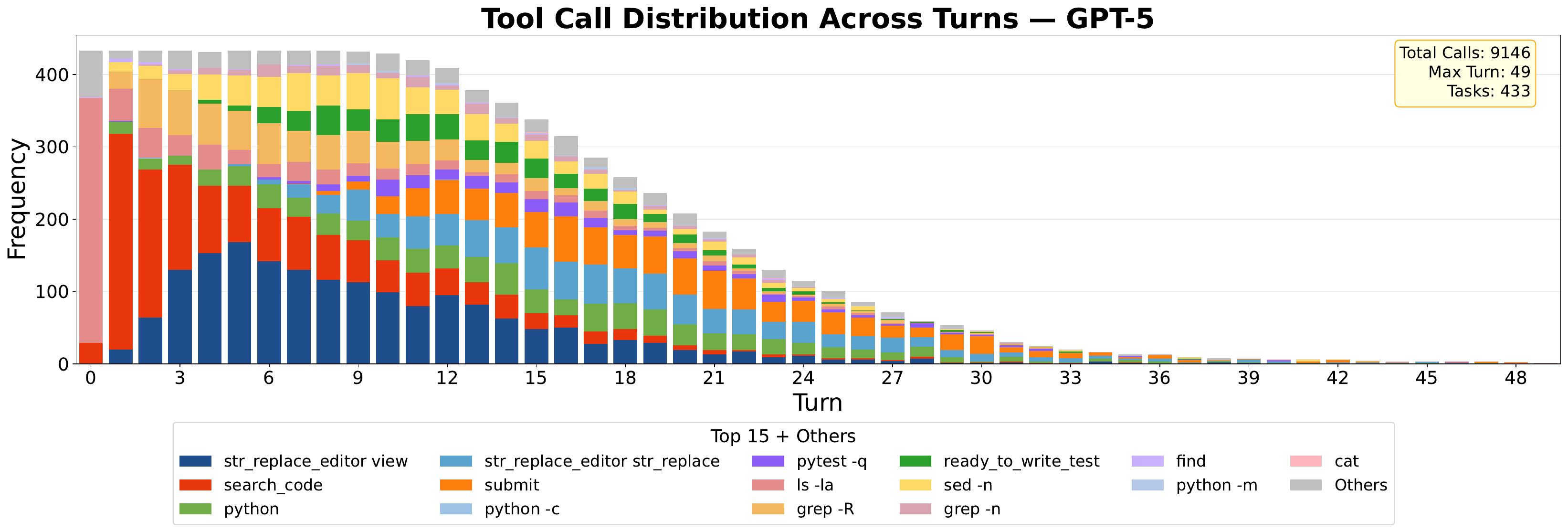}
    \caption{GPT-5}
    \label{fig:tool-gpt5}
\end{subfigure}\\[4pt]
\begin{subfigure}[t]{\textwidth}
    \centering
    \includegraphics[width=0.95\linewidth]{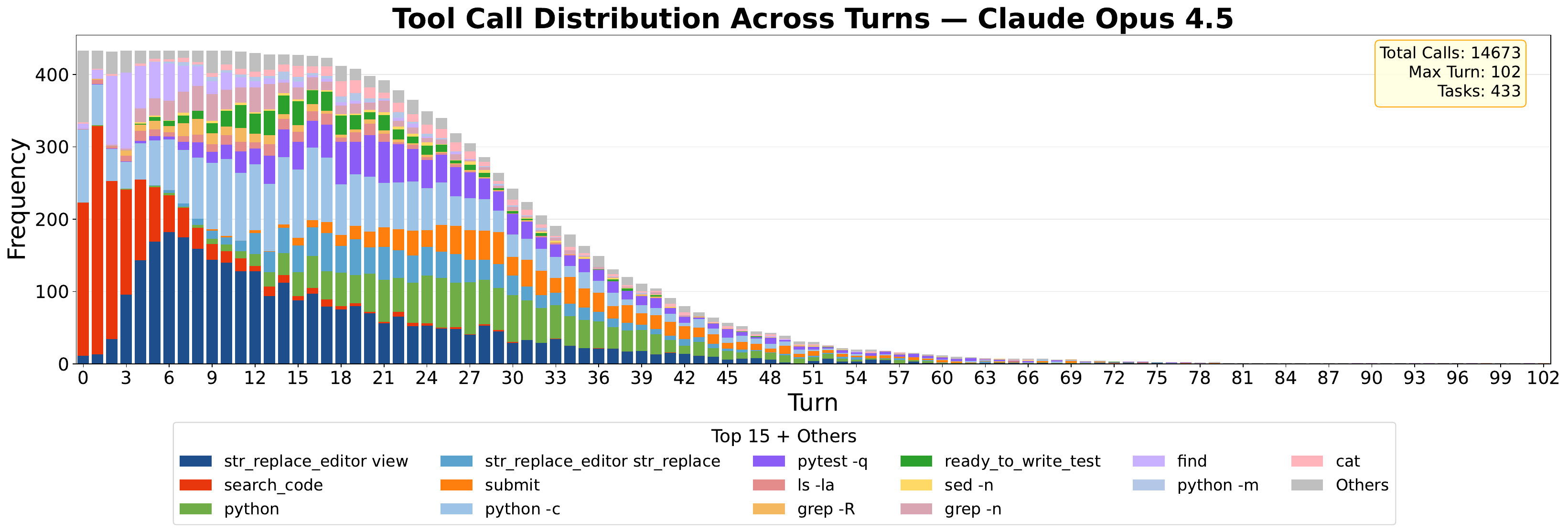}
    \caption{Claude-Opus-4.5}
    \label{fig:tool-opus}
\end{subfigure}
\caption{Tool call distribution across turns for three backbones.}
\label{fig:tool-distribution}
\end{figure}

\paragraph{Tool Usage Reveals a Two-Phase Workflow} To understand how our pipeline allocates exploration and generation effort, we visualize the tool call distribution across turns for all three backbones in Figure~\ref{fig:tool-distribution}. Two consistent patterns emerge across backbones.  First, \texttt{search\_code} dominates the early turns and disappears after \texttt{ready\_to\_write\_test} (typically around turns 8--16), reflecting our two-phase design that explicitly switches the tool set once exploration concludes; from that point on, the agent shifts to \texttt{str\_replace\_editor}  for focused test writing. Second, Python execution remains active throughout the trajectory, suggesting that the agent continuously validates its understanding via code execution in both phases. We further observe that stronger backbones such as GPT-5 transition to test writing earlier than GPT-5-Mini, reaching sufficient repository understanding with less back-and-forth.

\begin{figure}[t]
    \centering
    \begin{subfigure}[b]{0.32\textwidth}
        \includegraphics[width=\textwidth]{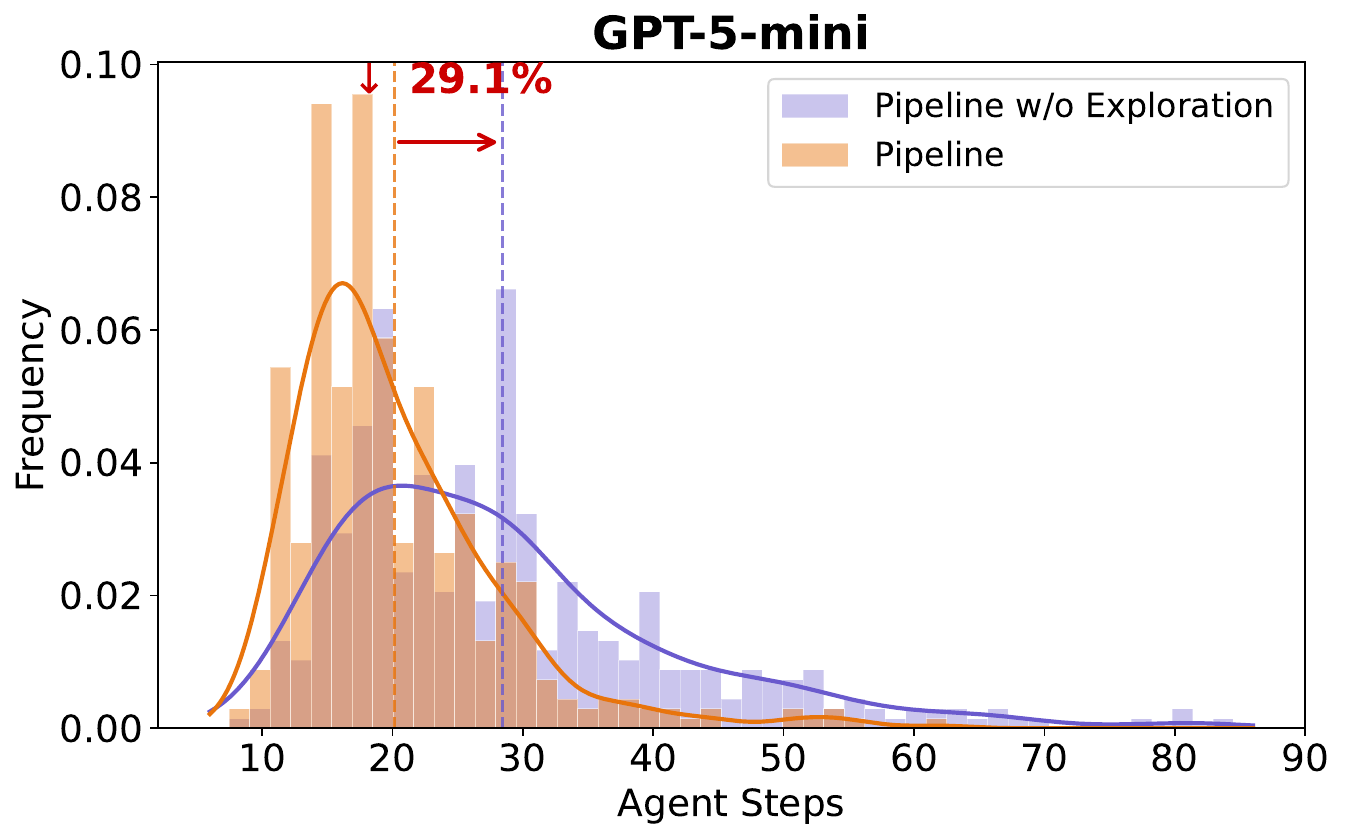}
        \caption{GPT-5-Mini}
    \end{subfigure}
    \hfill
    \begin{subfigure}[b]{0.32\textwidth}
        \includegraphics[width=\textwidth]{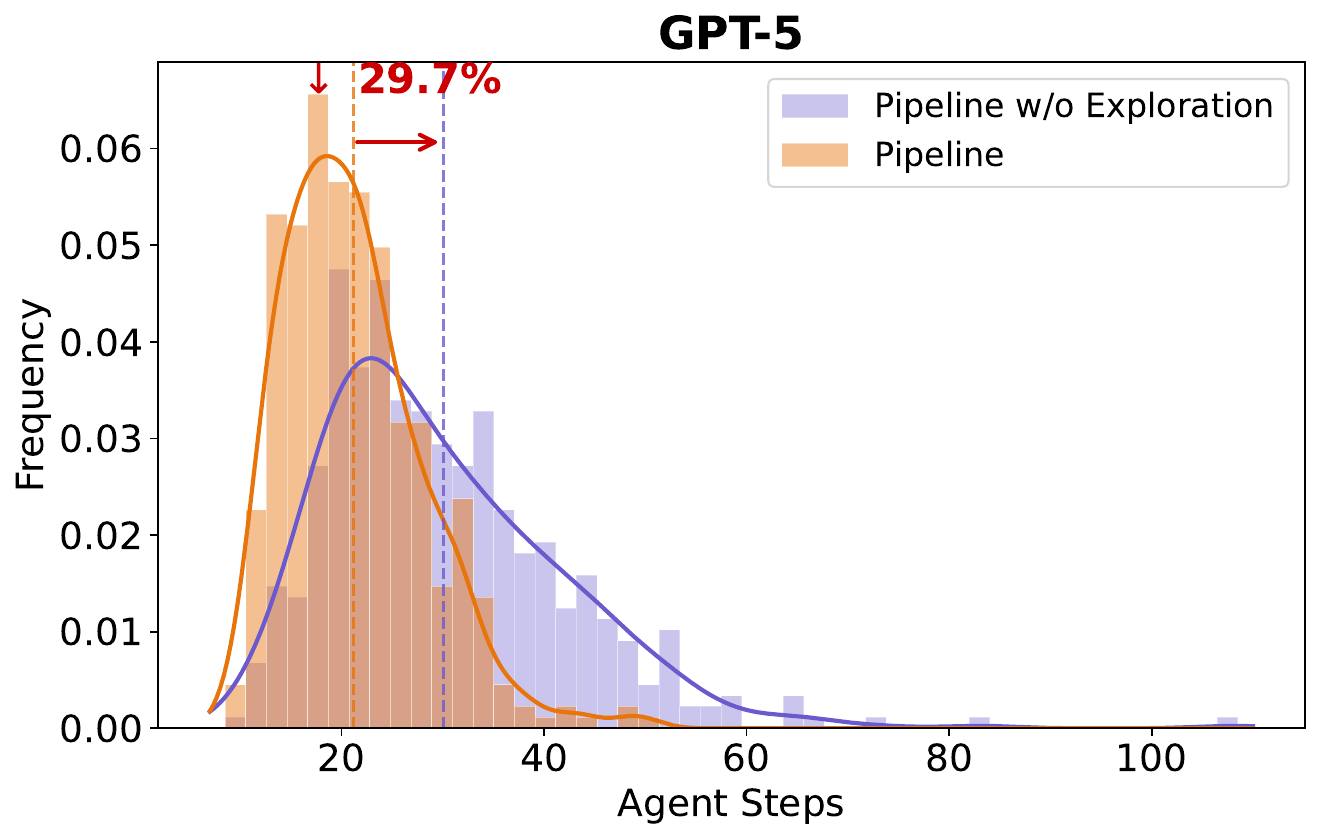}
        \caption{GPT-5}
    \end{subfigure}
    \hfill
    \begin{subfigure}[b]{0.32\textwidth}
        \includegraphics[width=\textwidth]{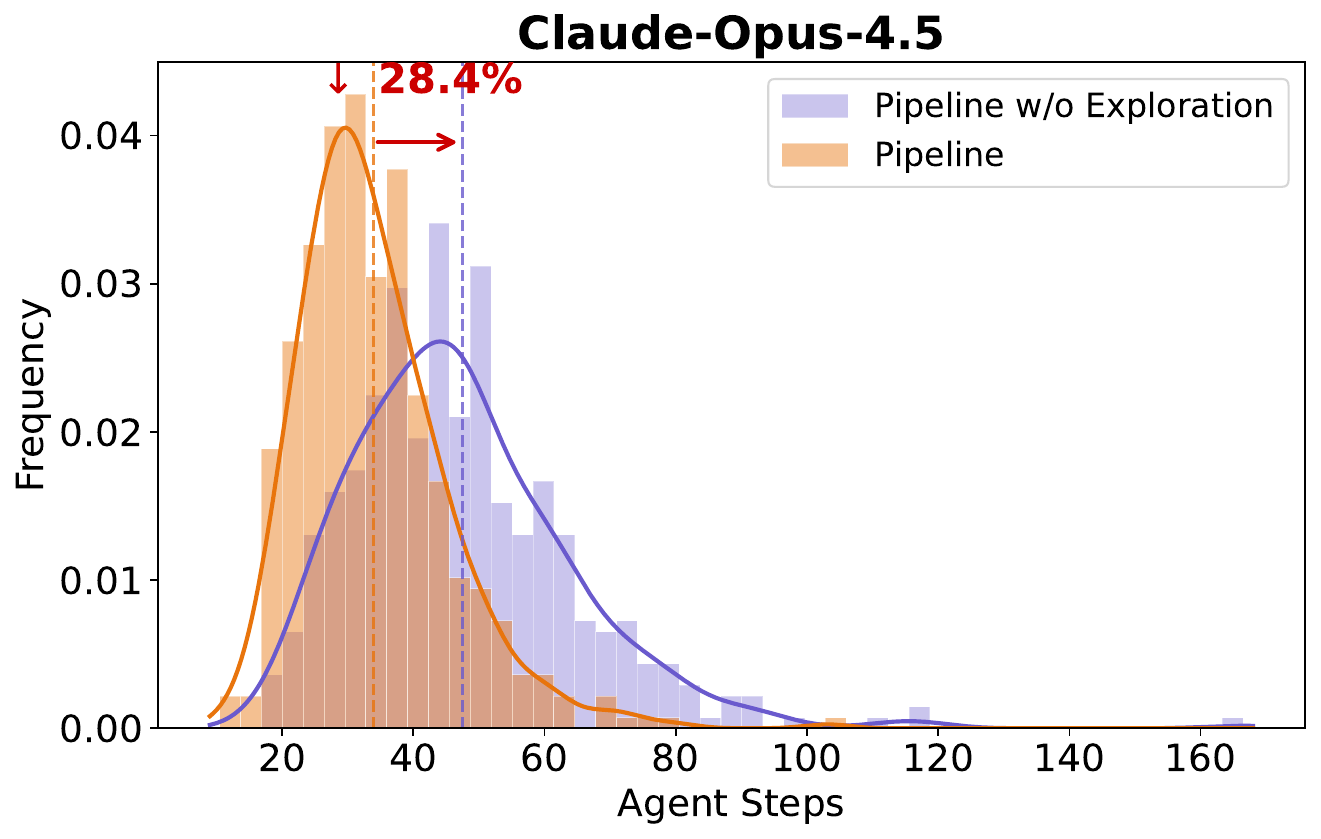}
        \caption{Claude-Opus-4.5}
    \end{subfigure}
    \caption{Distribution of agent steps for \mypipeline{} and the ablation without exploration across three backbones.}
    \label{fig:steps}
\end{figure}

\begin{figure}[t]
\centering
\begin{subfigure}[t]{0.32\textwidth}
    \centering
    \includegraphics[width=\linewidth]{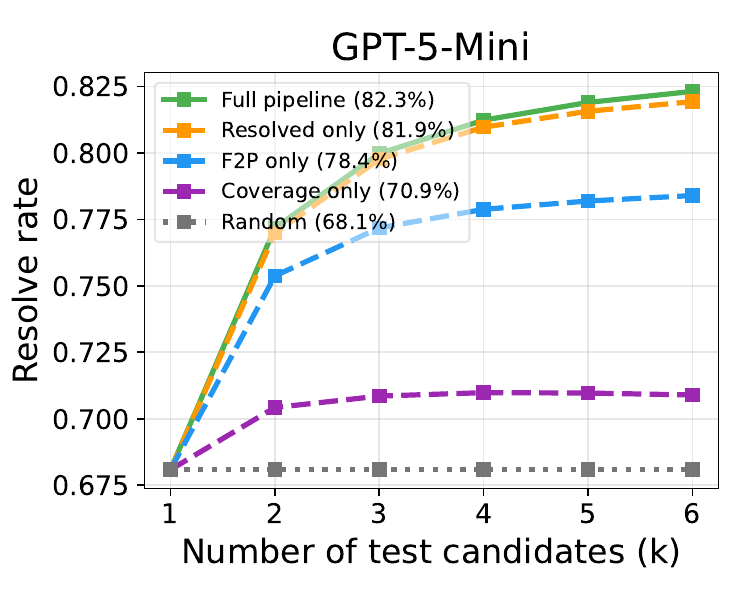}
    \caption{GPT-5-Mini}
    \label{fig:ablation-gpt5mini}
\end{subfigure}
\hfill
\begin{subfigure}[t]{0.32\textwidth}
    \centering
    \includegraphics[width=\linewidth]{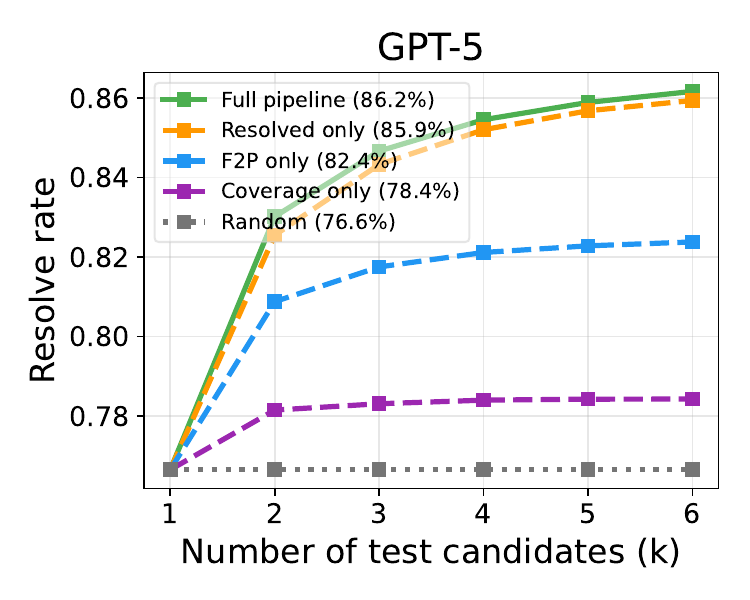}
    \caption{GPT-5}
    \label{fig:ablation-gpt5}
\end{subfigure}
\hfill
\begin{subfigure}[t]{0.32\textwidth}
    \centering
    \includegraphics[width=\linewidth]{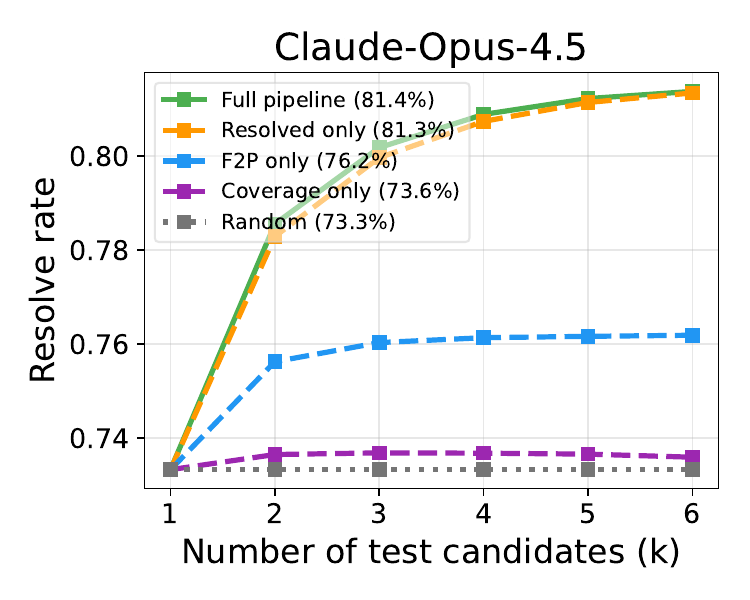}
    \caption{Claude-Opus-4.5}
    \label{fig:ablation-claude}
\end{subfigure}
\caption{Ablation of selection signals. Each variant uses $N$=3 code patches. The resolved count is the strongest individual signal, while coverage serves primarily as a tiebreaker.}
\label{fig:ablation-selection}
\end{figure}

\paragraph{Exploration Tools Reduces Total Agent Steps} Figure~\ref{fig:steps} shows the distribution of total agent steps for \mypipeline{} and the ablation without exploration across three backbones. 
Across all models, equipping the agent with exploration tools consistently shifts the step distribution leftward, with the median reduced by approximately 29\% compared to the variant without them (29.1\% for GPT-5-Mini, 29.7\% for GPT-5, and 28.4\% for Claude-Opus-4.5).
Without these tools, the agent must rely on primitive file browsing and keyword search to understand the repository, leading to more back-and-forth navigation steps before it can write meaningful tests. In contrast, our exploration tools provide structured, semantically-aware code retrieval, allowing the agent to locate relevant code efficiently and proceed to test writing with fewer steps. 

\paragraph{Ablation of Selection Signals} We ablate the contribution of each signal in our selection pipeline (Figure~\ref{fig:ablation-selection}), using $N$=3 code patches across all variants. The resolved count is the dominant signal, achieving within 0.3--0.4\% of the full pipeline across all three backbones. The F2P filter alone provides moderate gains over random (e.g., 10.3 points on GPT-5-Mini), serving as a coarse screen that eliminates candidates showing no interaction with the code patch. Coverage ranking alone adds only 2--3 points beyond random, indicating its primary role as a tiebreaker. The full pipeline's marginal improvement over resolved-only ranking comes from F2P pre-screening, which removes low-signal candidates and helps disambiguate ties among top-ranked tests.

\paragraph{Selected Test Source Distribution} Table~\ref{tab:test-source} reports the distribution of selected test patches across pipeline configurations. The \textit{dropCode} variant is the most frequently selected source for both GPT-5-Mini (26.6\%) and GPT-5 (24.3\%), suggesting that removing code context during test generation encourages more diverse and discriminative tests. For Claude-Opus-4.5, the full pipeline configuration dominates (29.4\%), likely because its stronger baseline generation already produces high-quality tests that benefit less from constrained prompting. No single source accounts for more than 30\% of selections across any backbone, confirming that diverse generation strategies contribute complementary test candidates and justifying the multi-configuration design.

\begin{table}[t]
\centering
\caption{Distribution of selected test patch sources (\%) after test selection ($k$=6, $N$=3). Bold indicates the most frequently selected source for each backbone.}
\label{tab:test-source}
\setlength{\tabcolsep}{4pt}
\renewcommand{\arraystretch}{1.35}
\resizebox{0.95\textwidth}{!}{%
\begin{tabular}{l R R R R R R}
\toprule
\textbf{Model} & \textbf{dropCode} & \textbf{initPatch} & \textbf{initTest} & \textbf{simple} & \textbf{standard} & \textbf{Pipeline} \\
\midrule
\colorbox{gpt5m}{\textcolor{gpt5mtext}{\texttt{\textbf{GPT-5-Mini}}}} & \textbf{26.6} & 21.1 & 18.3 & 12.0 & 10.0 & 12.0 \\
\colorbox{gpt5}{\textcolor{gpt5text}{\texttt{\textbf{GPT-5}}}} & \textbf{24.3} & 21.5 & 16.9 & 11.6 & 12.7 & 13.0 \\
\colorbox{claude45}{\textcolor{claude45text}{\texttt{\textbf{Claude-Opus-4.5}}}} & 20.4 & 16.4 & 13.9 & 10.2 & 9.7 & \textbf{29.4} \\
\midrule
\rowcolor{gray!8}
\textit{Average} & \textbf{23.8} & 19.7 & 16.4 & 11.3 & 10.8 & 18.1 \\
\bottomrule
\end{tabular}
}
\end{table}

\section{Limitations}
\label{limitation}
Our evaluation focuses on Python repositories from SWT-Bench; while the DPI principles are language-agnostic, validating on additional languages remains future work. Additionally, like all LLM-based approaches, inference cost scales with repository size, and further efficiency improvements are possible through more selective context management strategies.

\section{Broader impacts}
\label{Broader}
Our work focuses on automating reproduction test generation, which we believe carries a net positive societal impact by reducing developer burden and improving software reliability. We identify two potential risks and corresponding mitigations.

First, LM-generated tests may execute harmful operations during iterative verification (e.g., modifying or deleting files). To mitigate this, our pipeline runs all generated tests in isolated Docker containers, preventing generated code from affecting the host system.

Second, improved test generation could be misused to probe software for exploitable vulnerabilities. However, our system targets \emph{known} bug reports and requires an existing issue description, limiting offensive use. Releasing our framework can help the community develop safeguards for LM-based software engineering agents, while the defensive benefits of automated testing outweigh the risks.

\end{document}